\documentclass[prb,12pt, superscriptaddress]{revtex4}
\usepackage[T1]{fontenc}
\usepackage[latin9]{inputenc}
\usepackage{times}
\usepackage{units}
\usepackage{braket}
\usepackage{mathrsfs}
\usepackage{amsmath}
\usepackage{amssymb}
\usepackage{graphicx}
\usepackage{esint}
\usepackage{color}

\usepackage{morefloats}
\usepackage{times}



\usepackage{times}

\begin{document}

\title{
Modified Pad\'e approach for analytic continuation:\\
Application to the zero-gap Kondo lattice model
}

\author{Annam\'{a}ria~Kiss} \email{kiss.annamaria@wigner.mta.hu} 
\affiliation{Institute for Solid State Physics and Optics, Wigner Research Centre for Physics of the Hungarian Academy of Sciences, P. O. B. 49, H-1525, Budapest, Hungary}

\date{\today}
\begin{abstract} 

A modified Pad\'e approach is presented as analytic continuation for numerical methods that evaluate correlation functions in imaginary time. Instead of the direct analytic continuation of the correlation functions, the Pad\'e method is applied for the self-energy that is then used for deriving the Green's functions at real energies. We find that this modified, self-energy Pad\'e approach is more stable and robust against statistical errors compared to the direct way. The characteristics and success of the modified Pad\'e approach are analyzed by actual calculations for the illustrative cases of the non-interacting Anderson lattice and Hubbard model. A zero-gap Kondo lattice model with linearly vanishing conduction electron density of states at the Fermi level is also studied, where we use the modified Pad\'e approach as analytic continuation. We investigate the properties of the Kondo insulating state including the dependence of the insulating gap on the Kondo coupling and coherence effects. Furthermore, we identify two energy scales from dynamic and thermodynamic quantities, that are associated as a direct and an indirect gap in a band hybridization picture.

\end{abstract}
\maketitle

\section{Introduction}

The physics of strongly correlated electron systems is very rich due to the interplay and competition of different energy scales. In the case of rare-earth systems, the Kondo lattice model (KLM) is the simplest attempt to investigate heavy electrons as a consequence of collective Kondo effect.
The easiest way to study the KLM is the application of dynamical mean-field theory (DMFT), although it becomes exact only at infinite dimension.
Quantum Monte Carlo (QMC) approaches are essential computation methods in condensed matter physics, and are often used as impurity solver for DMFT calculations as well.

In QMC methods that evaluate imaginary time path integrals quantities like Green's functions and response functions are obtained in imaginary time.
In order to interpret physical quantities measured experimentally at real energies, the Matsubara frequency ($i\varepsilon_{n}$) data obtained by Fourier transform from the imaginary-time results have to be analytically continued to real energies, $\varepsilon$.
The subject of analytic continuation is an important issue in numerical physics, however, it is very difficult because of its ill-posed nature.
When the function ${\cal F}(i\varepsilon_{n})$ under question is a rational function of $i\varepsilon_{n}$, the replacement $i\varepsilon_{n} \rightarrow \varepsilon + i\delta$ can be done. However, it is not correct when ${\cal F}(i\varepsilon_{n})$ is not a rational function. 
Furthermore, from the ill-posed nature it follows that the output depends sensitively on the input, therefore, statistical errors that are always present in numerical data make the analytic continuation scheme instable.
The simplest and widely used approach to perform analytic continuation is the Pad\'e approximation\cite{Pade} where the real-energy function in question is approximated by a continued fraction using the Matsubara frequency data points.
Other methods are also often used for analytic continuation in numerical physics such as the Maximum Entropy method\cite{MaxEnt}.

Efficient QMC impurity solvers such as the continuous-time quantum Monte Carlo (CT-QMC)\cite{CTQMC} method  provide very accurate numerical results in imaginary time.
Therefore, the simple Pad\'e method for analytic continuation can be successfully applied for most impurity problems.
This widely accepted method is very popular because of its simplicity or easy programmability, and it is often used also for less efficient, conventional QMC methods.
However, it has disadvantages as well including the overestimation of statistical errors or cut-off effects due to the limited number of Matsubara data points.
Another problem would be that when we solve lattice problems by DMFT with considerable number of self-consistency steps, less accurate results are derived compared to impurity cases with a given amount of computational time even when we use efficient QMC methods.
Thus, the application of Pad\'e method as analytic continuation is often questionable.

In this paper we present a modified Pad\'e approach which is based on the analytic continuation of the self-energy instead of the Green's function. 
Namely, 
we apply the Pad\'e analytic continuation method for the conduction electron self-energy known at imaginary-frequency points, and then calculate the Green's functions and $t$-matrix at real energies from the analytically continued, real-energy self-energy.
We show that this modified, self-energy Pad\'e approach works better than the standard way based on the direct analytic continuation of the Green's function especially for systems that have bare conduction electron density of states (DOS) with singular behavior at band edges or around the Fermi level. 
We found that in Ref.~\onlinecite{Wang} while the effect of antiferromagnetic ordering was studied on the properties of a Mott insulator, the analytic continuation of the self-energy was proposed. 
However, the authors neither used the Pad\'e method for analytic continuation nor discussed the properties and efficiency of the self-energy analytic continuation. 
We discuss the reasons for the success of the modified Pad\'e approach including its robustness against statistical errors or stability by actual calculations for the non-interacting Anderson lattice model, and also for the Hubbard model.
Finally we apply the modified, self-energy Pad\'e method for the zero-gap Kondo lattice with linearly vanishing bare conduction electron DOS at the Fermi level as well, and discuss the properties of its Kondo insulating state.

The so-called pseudogap or zero-gap Kondo impurity models with vanishing conduction electron DOS at the Fermi level as $\rho(\omega) \sim |\omega|^{r}$ have been extensively studied\cite{Withoff,  Chen, Fritz_2004, Bulla1, Buxton, Vojta, Vojta_Fritz, Bulla_Glossop, Ingensent} in the last two decades.
Renormalization group calculations have given comprehensive analytical account for the fixed-point structure, phase transitions and critical properties\cite{Withoff, Chen, Fritz_2004, Buxton, Vojta_Fritz},
while numerical renormalization group (NRG) studies have provided numerical calculations of dynamic and thermodynamic quantities\cite{Ingensent, Bulla_Glossop, Bulla1, Vojta}. 
General understanding for the pseudogap Kondo impurity model involves a quantum phase transition at a critical Kondo coupling $J_{c}$ above which the impurity is Kondo screened while it is in a local moment state for $J<J_{c}$. 
For $r \ge 1/2$ the transition disappears for particle-hole symmetric case and the impurity is always unscreened independent of the value of Kondo coupling $J$. 
However, with particle-hole asymmetry the phase transition is present for all $r>0$.
In contrast to the impurity version of the zero-gap Kondo model, less is known about its lattice version.
The Dirac Kondo lattice model as the extension of the impurity pseudogap Kondo model with exponent $r=1$ has been studied by mean-field methods\cite{Principi, Zhong} and QMC simulations\cite{Liu, Fu}. 
Refs.~\onlinecite{Zhong, Liu, Fu} have investigated the phase diagram of the Dirac Kondo lattice with possible electronic orders by extended mean-field study\cite{Zhong} and QMC method\cite{Liu, Fu}, however, less attention has been paid to the Kondo insulating state. 
In our study on the zero-gap Kondo lattice with $r=1$ we concentrate on its insulating phase. This $r=1$ case might have relevance for the case of Kondo physics in graphene or Kondo correlation in $d$-wave superconductors\cite{Fritz}.
Based on the accurately derived real-energy Green's function and $t$-matrix obtained by the modified Pad\'e approach, we discuss dynamical and thermodynamical properties of the Kondo insulating phase including the optical conductivity, static local susceptibility, dependence of the insulating gap on the Kondo coupling $J$, and coherence effects.
We identify two energy scales as well from the dynamic and thermodynamic properties, that are associated as a direct and an indirect gap in a band hybridization picture.

\section{The modified analytic continuation method}\label{modified-analytic-continuation}

In QMC methods that evaluate imaginary-time path integrals the conduction electron Green's function is obtained in the imaginary time domain.
Interpretation of physical quantities measured experimentally requires the knowledge of Green's function at real energies, which is mathematically carried out by analytic continuation of the Matsubara data down to the real-energy axis by the procedure $i\varepsilon_{n}  \rightarrow  \varepsilon + i\delta$.

Numerically the conduction electron DOS, $\rho_{c}$, is ordinarily derived at real energies, $\varepsilon$, through the direct analytical continuation of the Matsubara Green's function $G_{c}(i\varepsilon_{n})$ as
\begin{eqnarray}
\rho_{c}(\varepsilon + i\delta) 
= -\frac{1}{\pi} {\rm Im}\, G_{c}(\varepsilon+i\delta),
\label{rhoc-gc1}
\end{eqnarray}
which will be addressed as Green's function approach, or direct analytic continuation method, in the following.

Instead of this direct way, there is a modified approach to obtain the spectral density $\rho_{c}(\varepsilon+i\delta)$ by using the analytically continued conduction electron self-energy, $\Sigma_{c}(\varepsilon + i\delta)$, derived from the Matsubara data $\Sigma_{c}(i\varepsilon_{n})$. 
Namely, the Green's function $G_{c}(\varepsilon +  i\delta)$ is expressed by the analytically continued self-energy at real energies as
\begin{eqnarray}
G_{c}(\varepsilon +  i\delta)  =   \int d\omega \rho(\omega) \left(\varepsilon + i\delta - \omega  + \mu - \Sigma_{c}(\varepsilon + i\delta) \right)^{-1}
\label{eq-greenc2} 
\end{eqnarray}
with $\rho(\omega)$ being the bare conduction electron DOS,
which gives $\rho_{c}$ as
\begin{eqnarray}
\rho_{c}(\varepsilon)  &=&  -\frac{1}{\pi}{\rm Im} \int d\omega  \rho(\omega) \left(\varepsilon + i\delta - \omega  + \mu - \Sigma_{c}(\varepsilon + i\delta) \right)^{-1}\nonumber\\
&=&
-\frac{1}{\pi}\int d\omega  \rho(\omega) 
\frac{{\rm Im} \, \Sigma_{c}(\varepsilon)}{(\varepsilon - \omega - {\rm Re} \, \Sigma_{c}(\varepsilon))^2 + ( {\rm Im} \, \Sigma_{c}(\varepsilon))^2}
\label{eq-greenc3} 
\end{eqnarray}
for ${\rm Im} \, \Sigma_{c}(\varepsilon) \gg \delta$, and we took $\mu = 0$ for simplicity.
Simplifying further the notation, we take the convention in this paper that energy including an infinitesimal imaginary part, $\varepsilon + i\delta$, is simply written as $\varepsilon$.
We will address the calculation of $G_{c}(\varepsilon)$ or $\rho_{c}(\varepsilon)$ from $\Sigma_{c}(\varepsilon)$ as self-energy approach, or modified analytic continuation method, in the following.

So far, both the Green's function and self-energy approaches were introduced generally. From now on we apply these two approaches with the Pad\'e analytic continuation method, whose details is given in Appendix~\ref{sec-app1}.
As a test for the modified Pad\'e approach we take the semicircular bare conduction electron DOS in this section given as
\begin{eqnarray}
\rho(\omega) = \frac{2}{\pi D^2}\sqrt{D^2-\omega^2}
\label{eq-semirho0}
\end{eqnarray}
with has sharp change at band edges, 
and examine the properties of the self-energy Pad\'e approach through the example of the non-interacting Anderson lattice and Hubbard model.

\subsection{Exemplary case of the non-interacting Anderson lattice model}

\subsubsection{Pad\'e approximation}

First we take the non-integrating Anderson lattice model (ALM) and discuss the main features of the Green's function and self-energy Pad\'e approaches.
Since this model can be solved exactly, we can compare the numerically obtained results\cite{note-ctqmc} with statistical errors to the exact ones.  
Namely, first we derive the 
local ($G_{f}$) and conduction electron ($G_{c}$) Green's functions  together with the self-energy $\Sigma_{c}$ numerically in the imaginary-time domain within DMFT. Then, we obtain the local and conduction electron spectral densities by the Green's function and self-energy Pad\'e analytic continuation approaches and compare them to the exact solutions.

The non-interacting ALM is given by the Hamiltonian
\begin{eqnarray}
{\cal H}_{\rm And, lattice} = \sum_{{\mathbf k},\sigma} \varepsilon_{{\mathbf k}}c_{{\mathbf k},\sigma}^{\dag}c_{{\mathbf k},\sigma} + \varepsilon_{f} \sum_{i,\sigma} f_{i,\sigma}^{\dag}f_{i,\sigma} 
+ V \sum_{i,\sigma} \left( f_{i,\sigma}^{\dag} c_{i,\sigma} +  c_{i,\sigma}^{\dag} f_{i,\sigma} \right),
\label{and-lattice-model}
\end{eqnarray}
where $f_{i,\sigma}$ ($f^{\dag}_{i,\sigma}$) and $c_{{\mathbf k},\sigma}$ ($c^{\dag}_{{\mathbf k},\sigma}$) are annihilation (creation) operators for the local and conduction electrons with the cite $i$ and spin $\sigma$ indices.
The dispersion $ \varepsilon_{{\mathbf k}}$ of the conduction electrons in Eq.~(\ref{and-lattice-model}) corresponds to the bare conduction electron DOS as $\rho(\omega) = 1/N_{0} \sum_{\mathbf k} \delta(\omega - \varepsilon_{{\mathbf k}})$ with $N_{0}$ being the number of lattice sites, where we take the semicircular conduction electron DOS given in Eq.~(\ref{eq-semirho0}) with $D=1$ in the numerical calculations. 
Analytic results for the non-interacting ALM that we will use in the following are summarized in Appendix~\ref{sec-app2}.

Figure~\ref{figure-anderson-1} shows the numerically calculated conduction electron and $f$-electron DOS at real energies derived by the Green's function Pad\'e approach ({\sl left panel}) and self-energy Pad\'e approach ({\sl right panel})  together with the exact solution.
It is clear that the real-energy results obtained by the self-energy Pad\'e approach show excellent agreement with the exact solution in spite of the presence of statistical errors, while in the case of the Green's function Pad\'e approach the obtained densities of states are less accurate.

\begin{figure}
\centering
\includegraphics[width= 0.45\hsize]{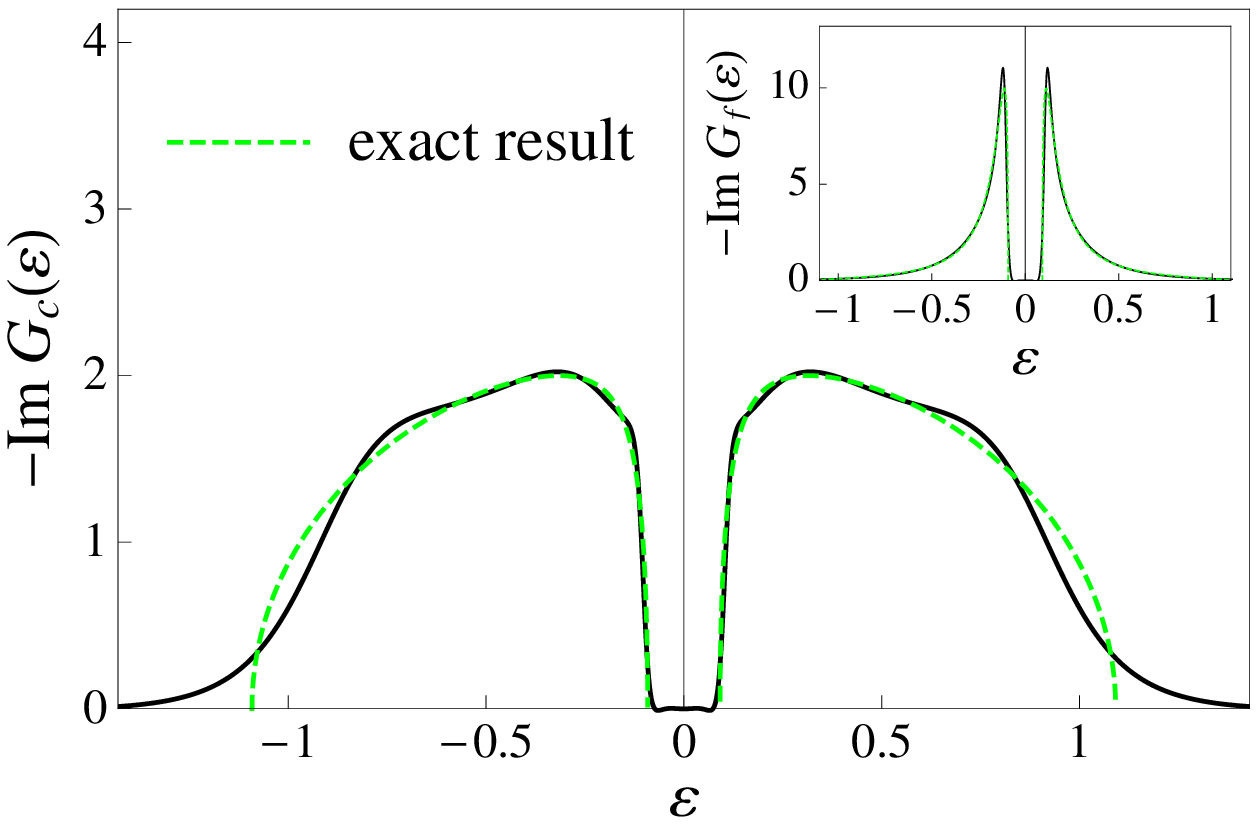}\hspace*{5mm}
\includegraphics[width= 0.45\hsize]{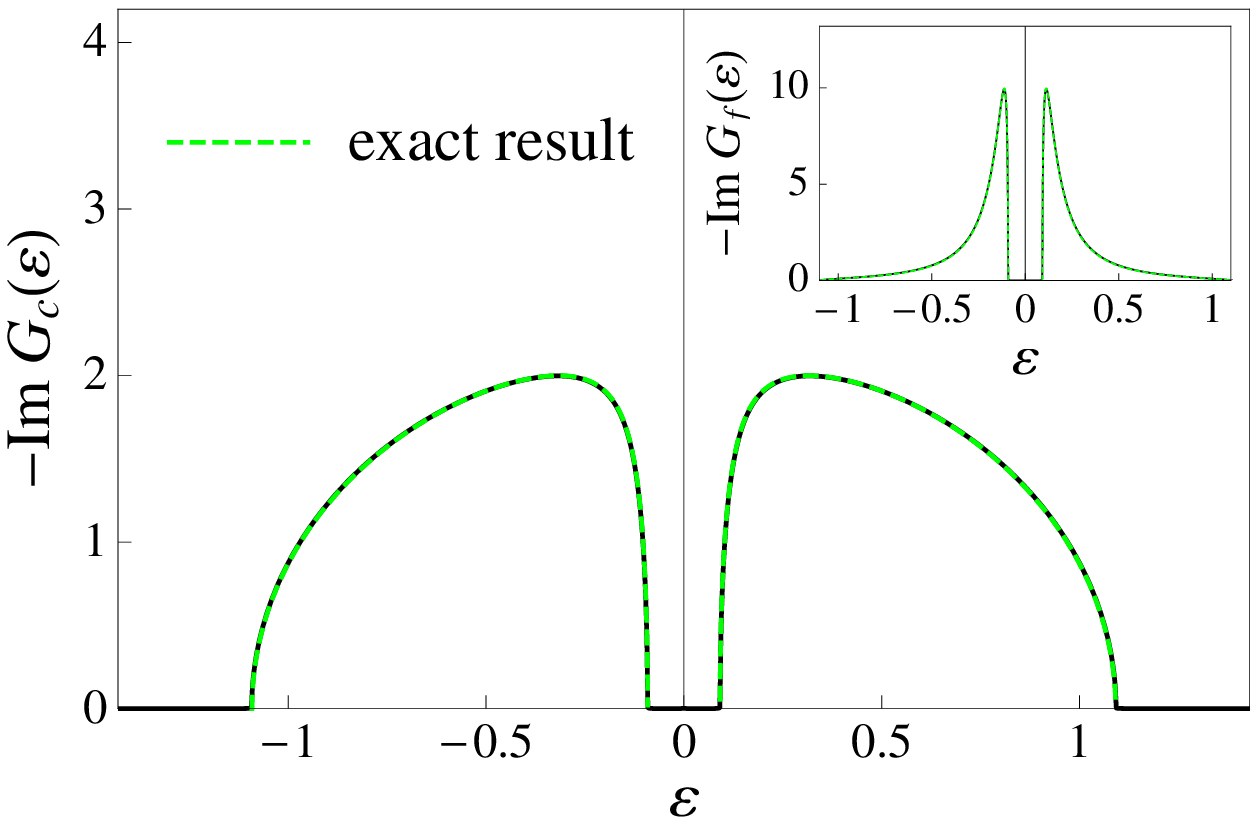}
\caption{Conduction electron ({\sl main}) and $f$-electron ({\sl inset}) DOS for the non-interacting Anderson lattice model obtained by the Green's function Pad\'e approach ({\sl left}), and by the self-energy Pad\'e approach ({\sl right}). {\sl Dashed green} curves show the exact result. The parameter values are chosen as $U=0$, $\varepsilon_{f}=0$, $\beta=500$, $V^2=0.1$, $D=1$.
}
\label{figure-anderson-1}
\end{figure}

Now we discuss the characteristics of the two Pad\'e approaches.
Considering the self-energy Pad\'e approach first, we start from the exact Matsubara self-energy  given as 
\begin{eqnarray}
\Sigma_{c}(i\varepsilon_{n}) = \frac{V^2}{i\varepsilon_{n}}
\label{Scz}
\end{eqnarray}
from Eq.~(\ref{eq-sigmac}) by taking $\varepsilon_{f}=0$, which is a rational function.
Therefore, it is immediately expected that the procedure $i\varepsilon_{n} \rightarrow \varepsilon + i\delta$ is correct and leads to the real-energy result
\begin{eqnarray}
\Sigma_{c}(\varepsilon) = \frac{V^2}{\varepsilon},
\label{Scz-ep}
\end{eqnarray}
which gives
\begin{eqnarray}
\rho_{c}(\varepsilon) = -\frac{1}{\pi} \frac{2}{D^2}{\rm Im}\, \left[ \varepsilon - \Sigma_{c}(\varepsilon) - i \sqrt{D^2 - (\varepsilon - \Sigma_{c}(\varepsilon))^2}  \right]
\label{eq-greenc4}, 
\end{eqnarray}
obtained after the $\omega$-integration in Eq.~(\ref{eq-greenc3}),
as 
\begin{eqnarray}
\rho_{c}(\varepsilon) = \frac{1}{\pi} \frac{2}{D^2}\sqrt{D^2 - \left(\varepsilon - \frac{V^2}{\varepsilon} \right)^2}
\label{eq-greenc4-II}, 
\end{eqnarray}
i.e. the exact result given in Eq.~(\ref{eq-rhoc-and-exact}) is recovered.
Actually, following the Pad\'e method described in Appendix~\ref{sec-app1}, the Pad\'e coefficients $a_{i}$ for the self-energy are calculated as
\begin{eqnarray}
a_{1} = \frac{V^2}{i\varepsilon_{0}}, \,\,\, a_{2} = -\frac{i}{\varepsilon_{0}}, \,\,\, a_{i} = 0 \,\,\, {\rm for}\,\,\, i>2
\end{eqnarray}
by taking  Eq.~(\ref{Scz}) for $\Sigma_{c}(i\varepsilon_{n})$,
which gives the real-energy Pad\'e approximant as
\begin{eqnarray}
\Sigma_{c}(\varepsilon) =  \frac{a_{1}}{1+ \frac{a_{2}(\varepsilon - i\varepsilon_{0})}{1+\frac{a_{3}(\varepsilon - i\varepsilon_{1})}{1+....}}  } =  \frac{a_{1}}{1+ a_{2}(\varepsilon - i\varepsilon_{0})  } = 
-i \frac{V^2}{\varepsilon_{0}} \frac{1}{[1  -\frac{i}{\varepsilon_{0}}(\varepsilon - i\varepsilon_{0})]} = \frac{V^2}{\varepsilon}
\label{eq-padapprSc}
\end{eqnarray}
as expected from Eq.~(\ref{Scz-ep}).
The rational function behavior of the self-energy guarantees the success of the modified Pad\'e approach as it was  demonstrated in the right part of Fig.~\ref{figure-anderson-1}.

\begin{figure}
\centering
\includegraphics[width= 0.45\hsize]{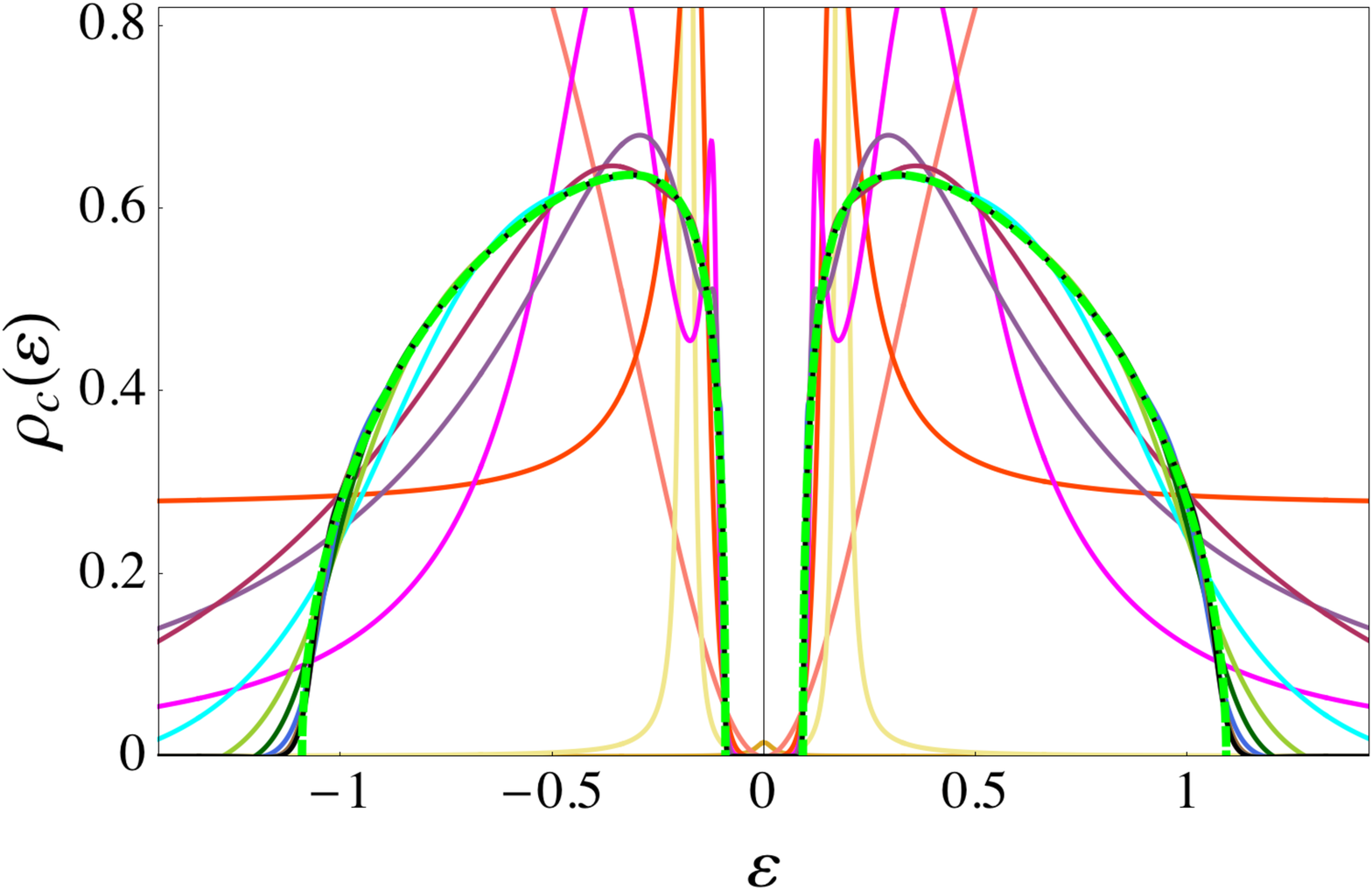}\hspace*{5mm}
\includegraphics[width= 0.45\hsize]{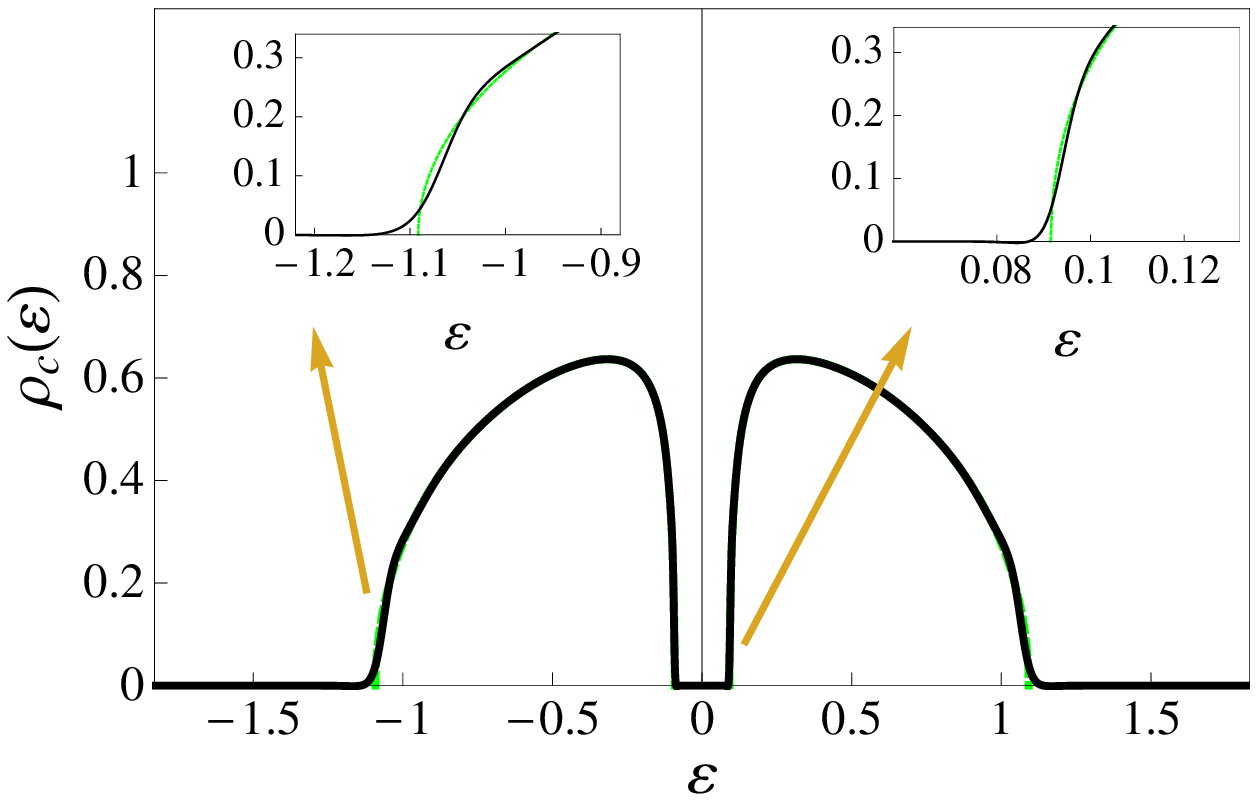}
\caption{{\sl Left:} Conduction electron DOS obtained from the Green's function Pad\'e approach with increasing number of the used Matsubara frequency points as $N=2, 3, 4, 5, 8, 10, 15, 20, 30, 50, 100, 200, 500, 1000$ colored from light to dark with increasing $N$. {\sl Right:} The converged result with $N=1000$. The {\sl insets} show enlarged view near sharp edges $|\varepsilon| \sim D$ and $|\varepsilon| \sim \varepsilon_{\rm F}$.
}
\label{figure-anderson-2}
\end{figure}

Considering next the Green's function Pad\'e approach, we start from the imaginary-frequency Green's function given as
\begin{eqnarray}
G_{c}(i\varepsilon_{n}) = \frac{2}{D^2} \left[ i\varepsilon_{n} - \frac{V^2}{i\varepsilon_{n}} - i \sqrt{D^2 - \left(i\varepsilon_{n} - \frac{V^2}{i\varepsilon_{n}} \right)^2}  \right]
\label{eq-Gcmatsanalytic}
\end{eqnarray}
from Eq.~(\ref{eq-greenc}) by taking Eqs.~(\ref{eq-semirho0}) and (\ref{Scz}), which is not a rational function. Thus, we cannot simply use the replacement $i\varepsilon_{n} \rightarrow \varepsilon + i\delta$, which implies that the exact result for $\rho_{c}(\varepsilon)$ given in Eq.~(\ref{eq-rhoc-and-exact}) will not be fully reproduced.
Indeed, when we consider the coefficients $a_{i}$ analytically for the Pad\'e approximant of the Green's function,
we find that all coefficients $a_{i}$ are non-zero in contrast to the case of the self-energy.
In left part of Fig.~\ref{figure-anderson-2} we show the real-energy Pad\'e approximants  for the Matsubara Green's function given in Eq.~(\ref{eq-Gcmatsanalytic}) with increasing number of Matsubara frequency points $N$ used in the calculation.
As we increase the number $N$, which is the only adjustable parameter in the Pad\'e method, the exact solution is approximated more and more.
However, we find that for $N$ larger than about $1000$ the result does not change apparently, and this converged Pad\'e result shown separately in the right part of Fig.~\ref{figure-anderson-2} differs from the exact result, especially at the sharp band and gap edges.

\subsubsection{Effect of statistical errors}

So far we did not considered the presence of statistical noises in the analytic calculations, however,
numerical calculations always contain statistical errors which are typically overestimated in the real-energy Pad\'e approximant giving rise to inaccuracy in the Pad\'e method.
In order to examine this effect, we 
model the presence of statistical noises in the conduction electron self-energy and Green's function by introducing random deviations from the exact results $\Sigma_{c}(i\varepsilon_{n})$ and $G_{c}(i\varepsilon_{n})$ for the ALM given in Eqs.~(\ref{Scz}) and (\ref{eq-Gcmatsanalytic}), respectively, at each Matsubara frequency point as
\begin{eqnarray}
\Sigma_{c}^{(\rm error)}(i\varepsilon_{n}) &=& \Sigma_{c}(i\varepsilon_{n}) + iq\delta_{r},
\label{Scerr}\\
G_{c}^{(\rm error)}(i\varepsilon_{n}) &=& G_{c}(i\varepsilon_{n}) + iq\delta_{r}.
\label{Gcerr}
\end{eqnarray}
Here, $\delta_{r}$ is a random number in the range $\{-1,1\}$ and $q$ is a control parameter which determines the "largeness" of the noise (error) in this toy model. 
In Eqs.~(\ref{Scerr}) and (\ref{Gcerr})
we add error only to the imaginary part of $\Sigma_{c}$ and  $G_{c}$ because they are pure imaginary due to the presence of particle-hole symmetry.

\begin{figure}
\centering
\includegraphics[width= 0.98\hsize]{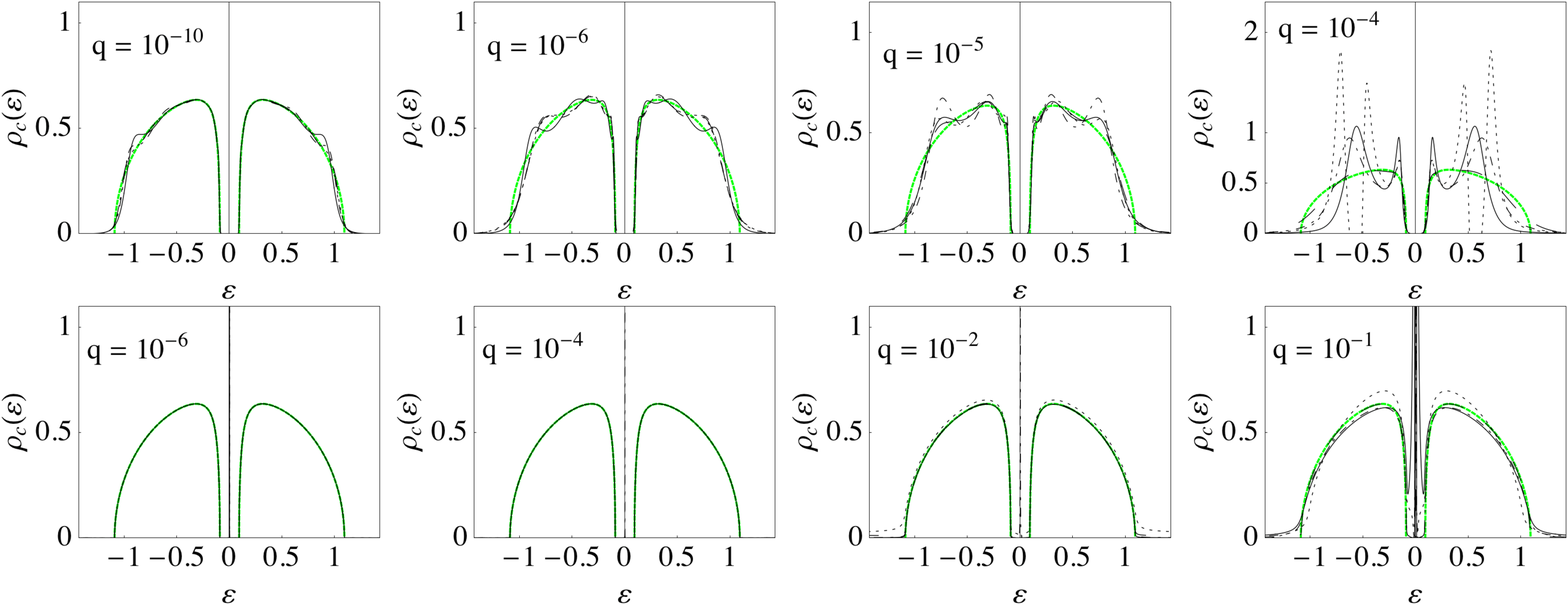}
\caption{Conduction electron DOS obtained by the Green's function Pad\'e ({\sl upper panel}) and self-energy Pad\'e ({\sl lower}) methods with different largeness of the error parameter $q$ defined in the main text. Each panel includes four different runs.
}
\label{figure-anderson-3}
\end{figure}

We take the functions $\Sigma_{c}^{(\rm error)}(i\varepsilon_{n})$ and $G_{c}^{(\rm error)}(i\varepsilon_{n})$ with noises and derive the analytically continued self-energy and Green's function by the Pad\'e method. Figure~\ref{figure-anderson-3} shows the spectral densities calculated from $\Sigma_{c}^{(\rm error)}(\varepsilon)$ and $G_{c}^{(\rm error)}(\varepsilon)$ for different values of the error parameter $q$.
We find that increasing the largeness of the random noises by increasing the value of parameter $q$ acts differently on the Green's function and self-energy Pad\'e approaches.
Namely, the self-energy Pad\'e approach is less sensitive to the  error compared to the Green's function approach, and remains stable even for "large" errors.
On the contrary, even small error causes large instability of the Pad\'e approximant in the Green's function method as it can be seen in the upper panel of Fig.~\ref{figure-anderson-3}. 
This toy model with $q \sim 10^{-6}$ depicts a situation similar to the numerical solution of the ALM shown in Fig.~\ref{figure-anderson-1}.

\subsection{Application of the modified Pad\'e method to the Hubbard model}\label{sec-hubbard}

In this Section we take another example which is the half-filled Hubbard model given by the Hamiltonian
\begin{eqnarray}
{\cal H}_{\rm Hubbard} = t \sum_{\langle i,j \rangle} c_{i}^{\dag}c_{j} + U \sum_{i} n_{i\uparrow} n_{i\downarrow},
\label{Hubb-model}
\end{eqnarray}
and obtain  the spectral density $\rho_{c}(\varepsilon)$ by both Pad\'e approaches numerically\cite{note-ctqmc} within DMFT. The bandwidth is $D=2t$ in Eq.~(\ref{Hubb-model}), and we take the semicircular bare conduction electron DOS given in Eq.~(\ref{eq-semirho0}).

\begin{figure}
\centering
\includegraphics[width= 0.3\hsize]{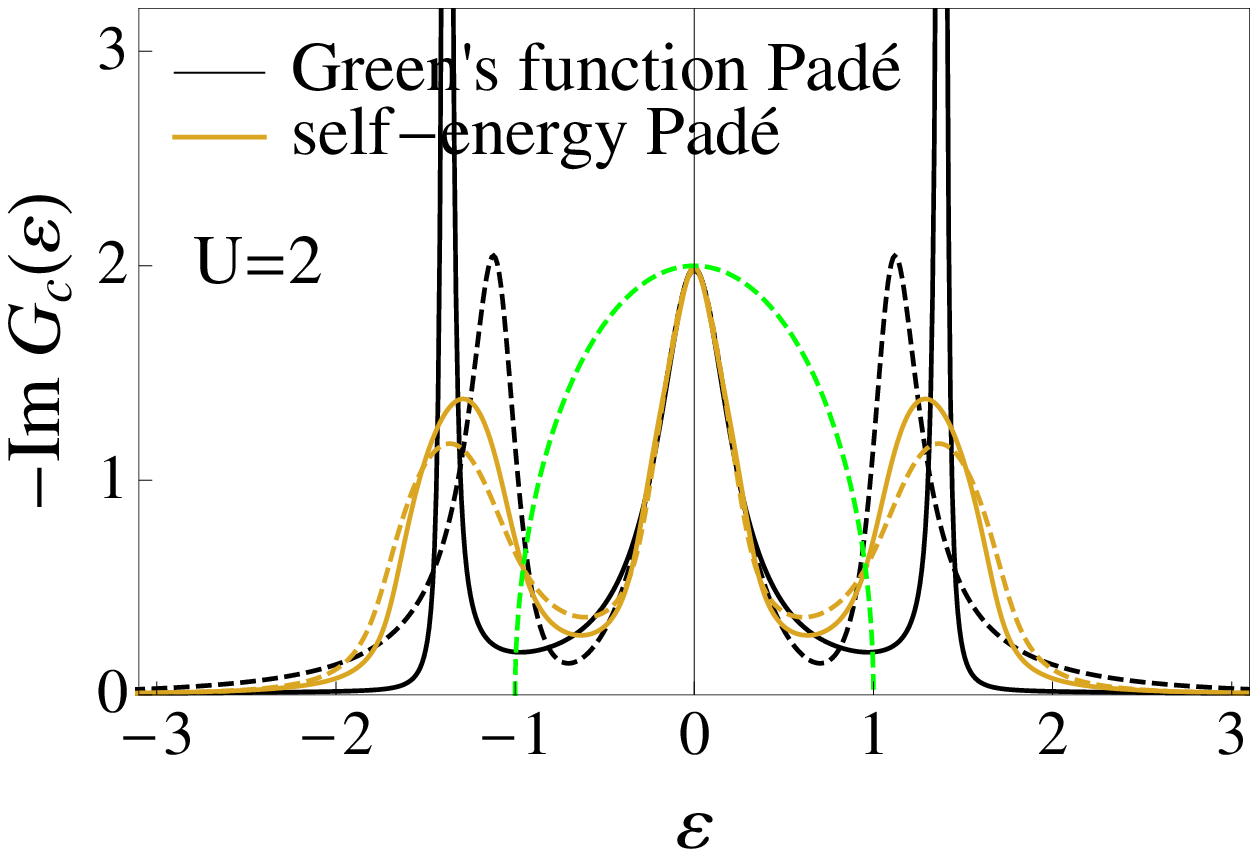}\hspace*{3mm}
\includegraphics[width= 0.3\hsize]{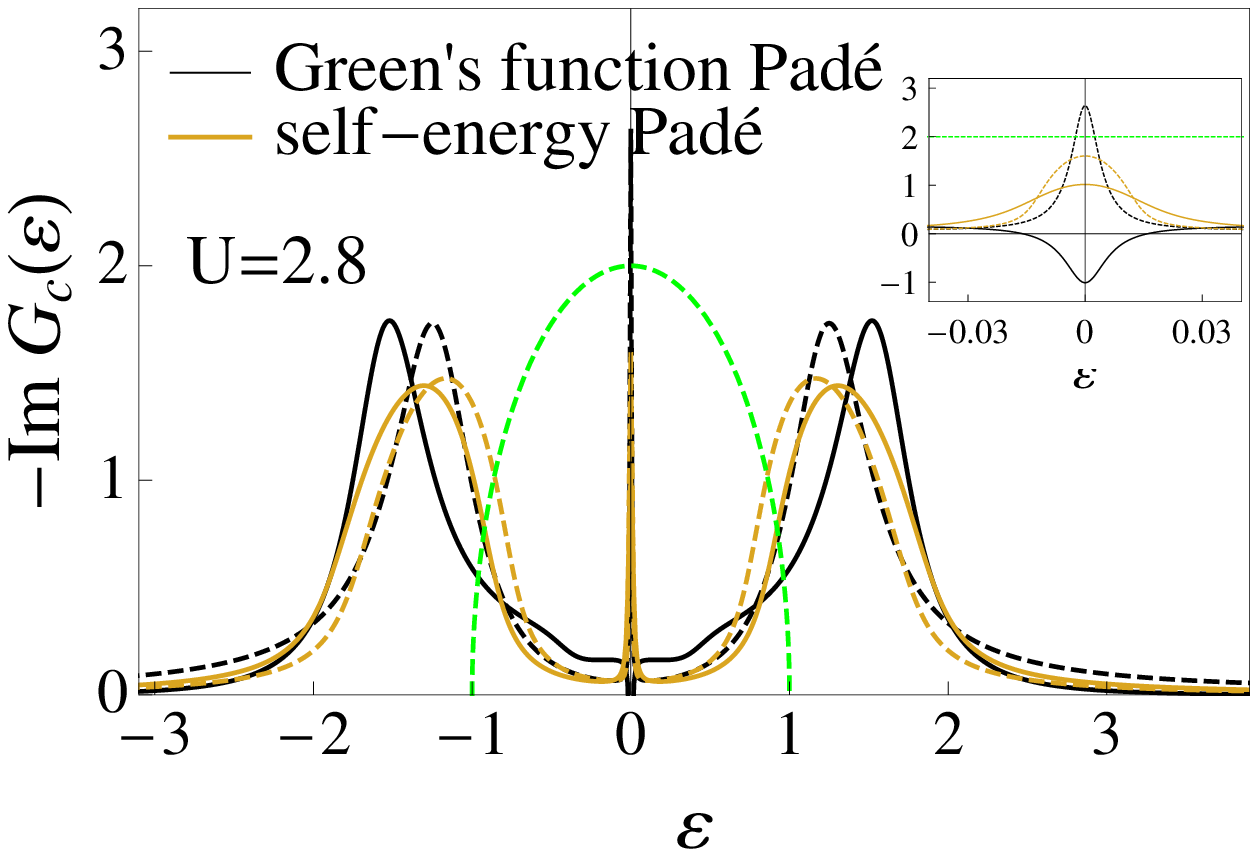}\hspace*{3mm}
\includegraphics[width= 0.3\hsize]{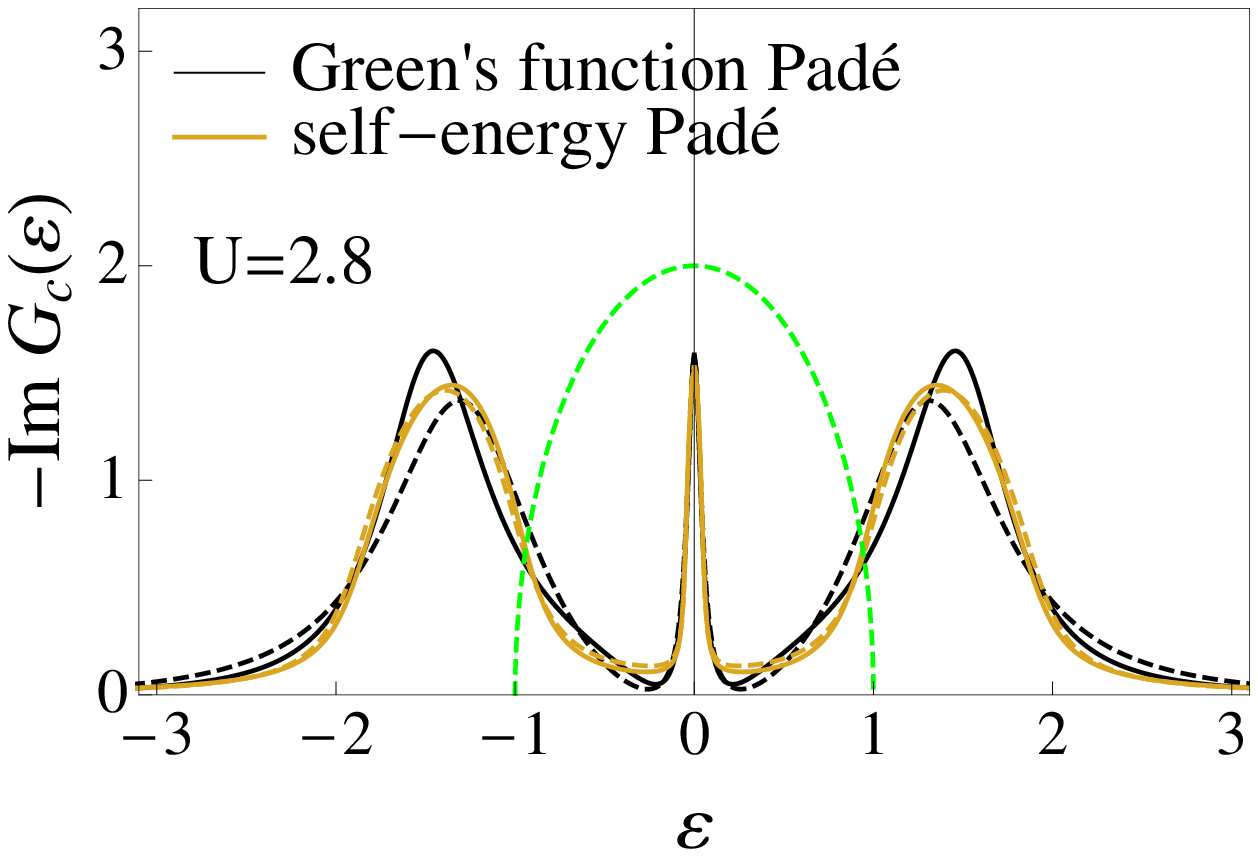}
\caption{Spectral densities at real energies obtained by the Green's function and self-energy Pad\'e approaches for $U=2$ ({\sl left}) and $U=2.8$ ({\sl center}). {\sl Right part} shows result for $U=2.8$ with larger number of Monte Carlo steps compared to the {\sl center panel}. The parameter values are chosen as $t=0.5$, $\beta=100$.
{\sl Continuous and dashed lines} correspond to spin-up and spin-down components, respectively, while the {\sl inset} in the {\sl center panel} shows the enlarged view of the spectrum around the Fermi level.
}
\label{figure-pade-hubbard-Gc}
\end{figure}

\begin{figure}
\centering
\includegraphics[width= 0.45\hsize]{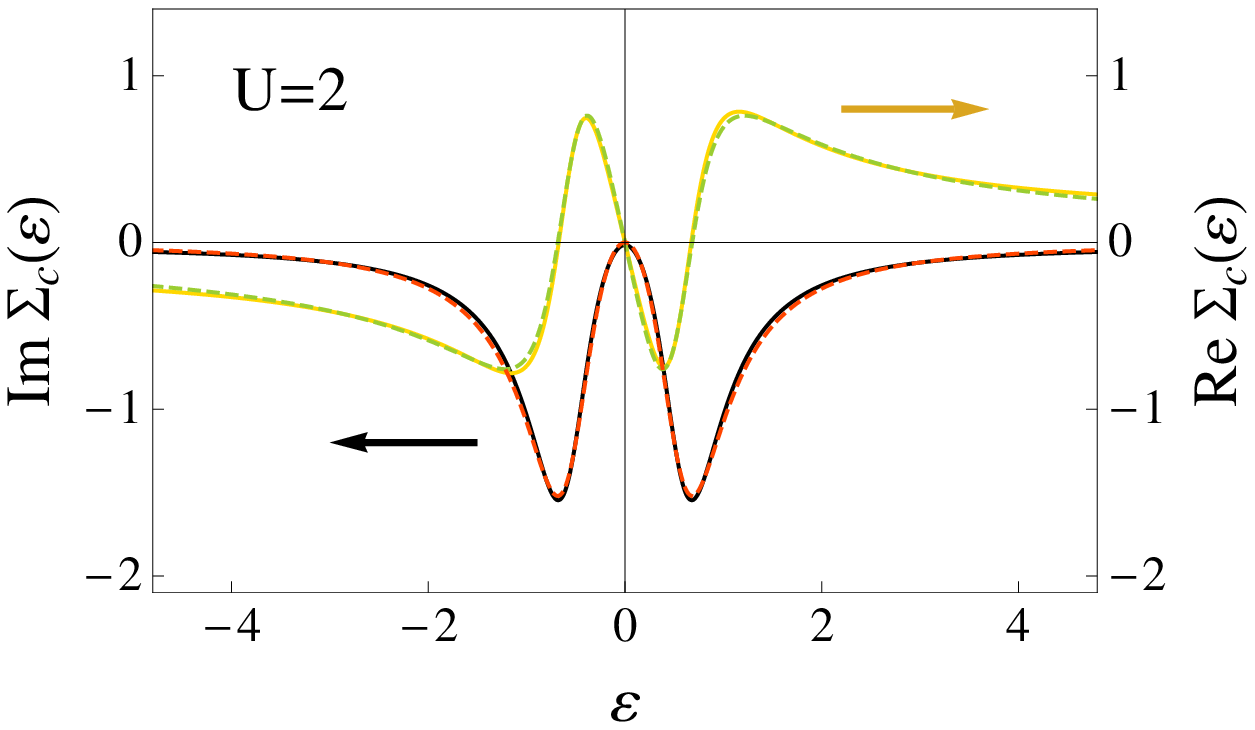}\hspace*{5mm}
\includegraphics[width= 0.45\hsize]{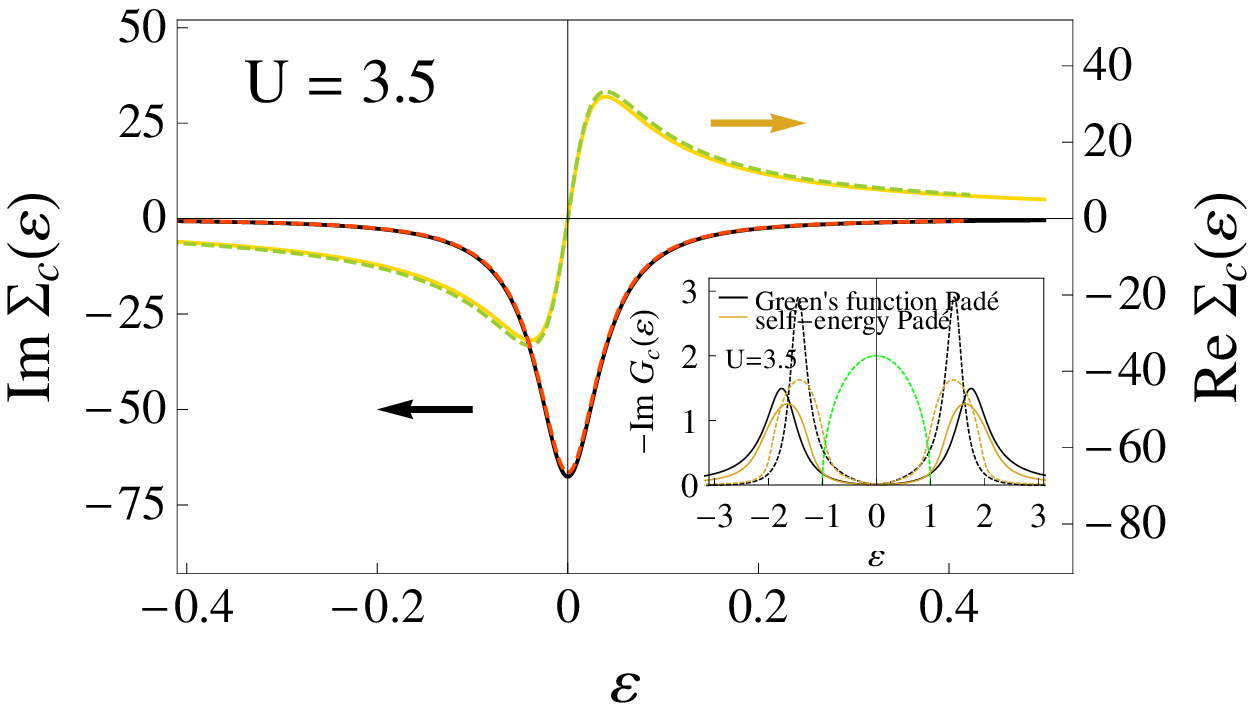}
\caption{Real-energy dependence of the self-energy for the Hubbard model at temperature $\beta=100$ with $U=2$ ({\sl left}) and $U=3.5$ ({\sl right}) choosing $D=1$ and $t=0.5$. {\sl Green and red dashed lines} are fitting results discussed in the main text.
The {\sl inset} shows the conduction electron DOS at real energies obtained by the Green's function and self-energy Pad\'e approaches with $U=3.5$.
}
\label{figure-selfenergy-hubbard-1}
\end{figure}

Figure~\ref{figure-pade-hubbard-Gc} shows the spectral densities as a function of real energy for Coulomb interaction values $U=2$ and $U=2.8$ obtained by both Pad\'e methods, while result for $U=3.5$ is shown in the inset of Fig.~\ref{figure-selfenergy-hubbard-1}.
We intentionally chose relatively small number of Monte Carlo steps in the simulations in the {\sl left} and {\sl center panels} of Fig.~\ref{figure-pade-hubbard-Gc} in order to test the accuracy of the two Pad\'e approaches.
In the right panel of Fig.~\ref{figure-pade-hubbard-Gc}, on the other hand, we show result with larger number of Monte Carlo steps compared to the {\sl center panel} as comparison.
For $U=2$ the ground state is a correlated metal which is indicated by a quasiparticle peak at the Fermi level. 
As we increase the value of $U$, the quasiparticle peak narrows, as for the case of $U=2.8$ in Fig.~\ref{figure-pade-hubbard-Gc}, and finally disappears in the insulating phase\cite{note-3} for which an example is shown in Fig.~\ref{figure-selfenergy-hubbard-1} with $U=3.5$.
The position of the metal-insulator transition is estimated as $U_{\rm cr}/D = 2.6$  by iterated perturbation theory\cite{dmft}, while it is found as  $U_{\rm cr}/D =2.94$ from DMFT calculation solved by NRG\cite{Bulla_Hubbard_NRG} for the Bethe lattice (semicircular DOS). 
We find $U_{\rm cr}/D = 2.9 \pm 0.05$ in our numerical calculations from the sharp change in the wave-function renormalization factor as a function of $U$, which coincides with the disappearance of the quasiparticle peak in the spectra.

The discrepancy between the results for the spin-up and spin-down components shown by dashed and continuous lines in Fig.~\ref{figure-pade-hubbard-Gc} mainly reflects  the presence of statistical errors.
We find again that the self-energy Pad\'e approach is less sensitive for the statistical errors while they cause large instabilities in the case of the Green's function Pad\'e approach.
Moreover, the Green's function approach gives even unphysical negative value for the spectral density as it can be seen in the inset of the center panel of Fig.~\ref{figure-pade-hubbard-Gc}.
When the number of the Monte Carlo steps is increased, the difference between the results obtained by the two Pad\'e approaches reduces.

To investigate the success of the self-enegy Pad\'e approach for the Hubbard model, we examine the properties of the conduction electron self-energy.
Fig.~\ref{figure-selfenergy-hubbard-1} shows the analytically continued self-energy both in the correlated metallic ($U=2$) and insulating ($U=3.5$) phases. 
Since the transition between these two phases is first-order, the self-energy has different shape in the two phases.
We find that the self-energy shows smooth energy dependence in both phases and can be described well phenomenolocically.
Namely, the phenomenological form
\begin{eqnarray}
\Sigma^{(\rm approx)}_{c, {\rm low-U}}(\varepsilon) = \sigma(\delta, z_1, y_1, \varepsilon)
\label{eq-selfenergy-approx-hubbard-I}
\end{eqnarray} 
describes the self-energy
in the correlated metallic phase with two-peak structure in its imaginary part,
while formula 
\begin{eqnarray}
\Sigma^{(\rm approx)}_{c, {\rm high-U}}(\varepsilon) = \sigma(0, 1, 1, \varepsilon) 
\label{eq-selfenergy-approx-hubbard-II}
\end{eqnarray}
describes it in the insulating phase with a one-peak structure in the imaginary part,
where the function $\sigma$ is defined as
\begin{eqnarray}
\sigma(\delta, z_1, y_1,\varepsilon) \equiv \frac{ -i \varepsilon z_{1} \Gamma}{\varepsilon y_{1}\Delta-i\varepsilon^2+i\delta}.
\label{eq-sigmafv}
\end{eqnarray}
The fits of the numerical data with these  formulas are also shown in Fig.~\ref{figure-selfenergy-hubbard-1}.

Since the $\varepsilon$-dependence of the self-energy can be excellently approximated by rational function in both phases, i.e. with formulas~(\ref{eq-selfenergy-approx-hubbard-I}), (\ref{eq-selfenergy-approx-hubbard-II}), it is expected that the Pad\'e approximant for the self-energy obtained from the Matsubara numerical data will be accurate.
Actually, taking the Matsubara phenomenological forms
\begin{eqnarray}
\Sigma^{(\rm approx)}_{c, {\rm low-U}}(i\varepsilon_{n}) &=& \sigma(\delta, z_1, y_1, i\varepsilon_{n}) =  \frac{ \varepsilon_n z_1\Gamma_1}{i \varepsilon_n y_1\Delta  + i (\varepsilon_n)^2+i\delta},
\label{eq-selfenergy-approx-hubbard-I-matsu}\\
\Sigma^{(\rm approx)}_{c, {\rm high-U}}(i\varepsilon_{n}) &=& \sigma(0, 1, 1, i\varepsilon_{n}) = \frac{\Gamma}{i\varepsilon_n + i\Delta}
\label{eq-selfenergy-approx-hubbard-II-matsu}
\end{eqnarray}
from Eqs.~(\ref{eq-selfenergy-approx-hubbard-I}), (\ref{eq-selfenergy-approx-hubbard-II}) by the inverse of the replacement $i\varepsilon_{n} \rightarrow \varepsilon + i\delta$,
the mathematically calculated real-energy Pad\'e approximants quickly converge and fully recover the real-energy phenomenological forms~(\ref{eq-selfenergy-approx-hubbard-I}) and (\ref{eq-selfenergy-approx-hubbard-II}) as it is shown in Appendix~\ref{app-hubbard}.

\section{Zero-gap Kondo lattice model}\label{zero-gap-Kondo}

\begin{figure}
\centering
\includegraphics[width= 0.3\hsize]{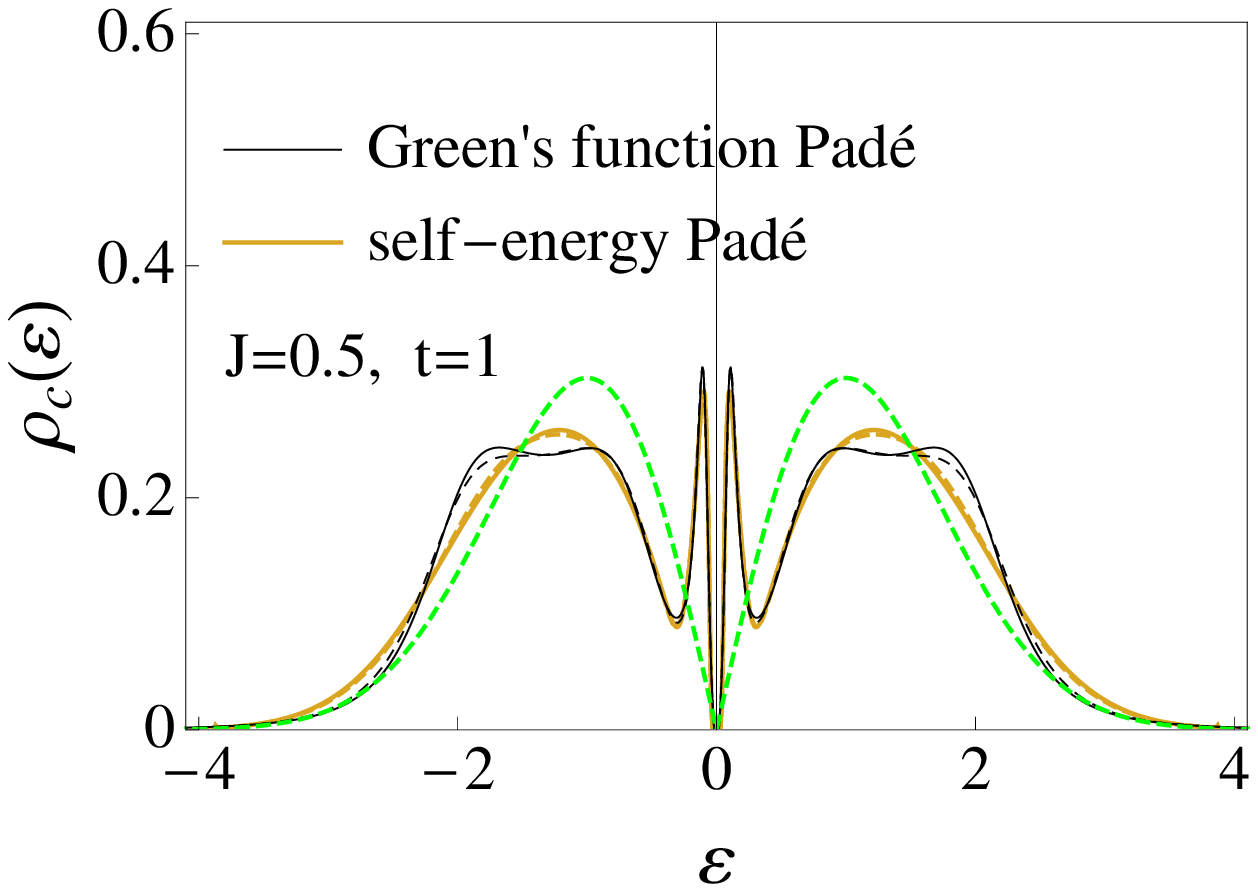}\hspace*{3mm}
\includegraphics[width= 0.3\hsize]{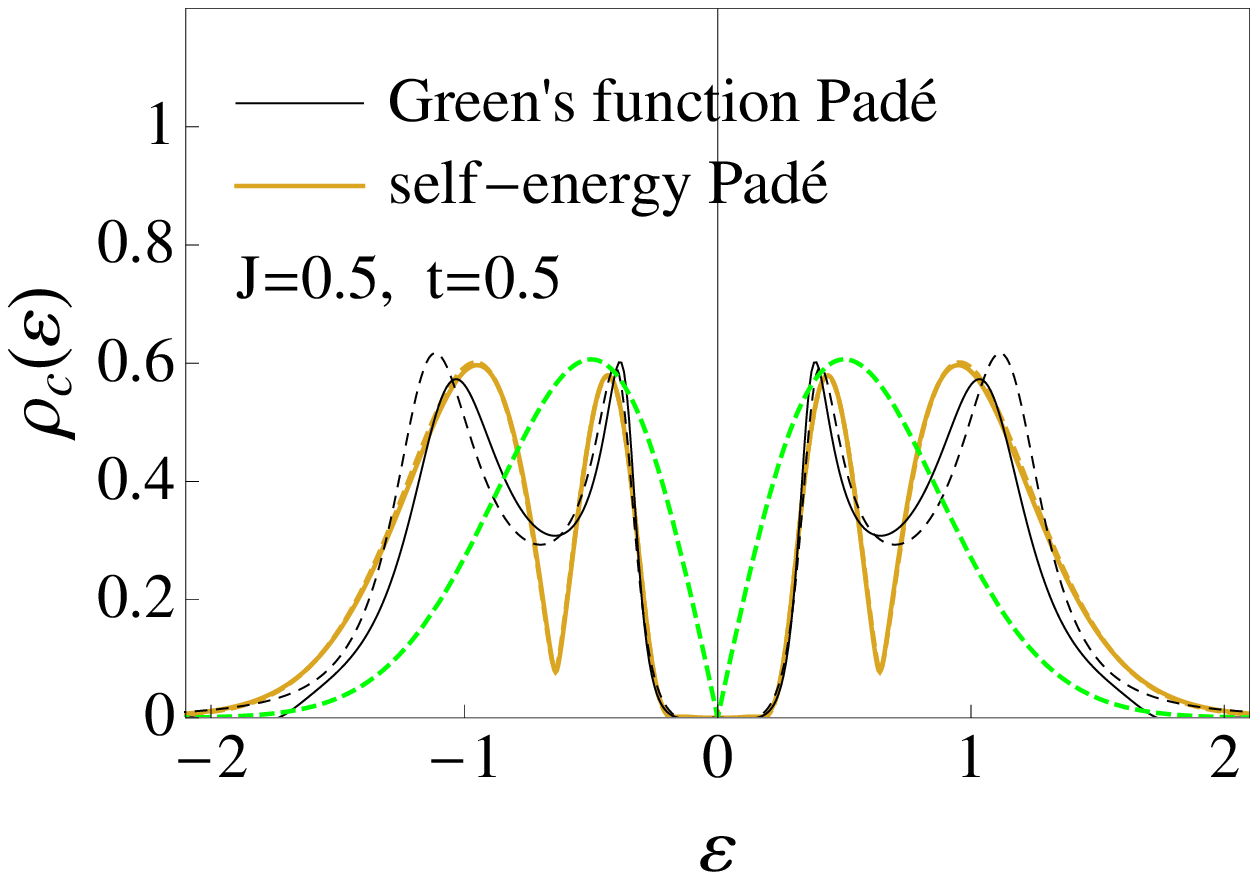}\hspace*{3mm}
\includegraphics[width= 0.3\hsize]{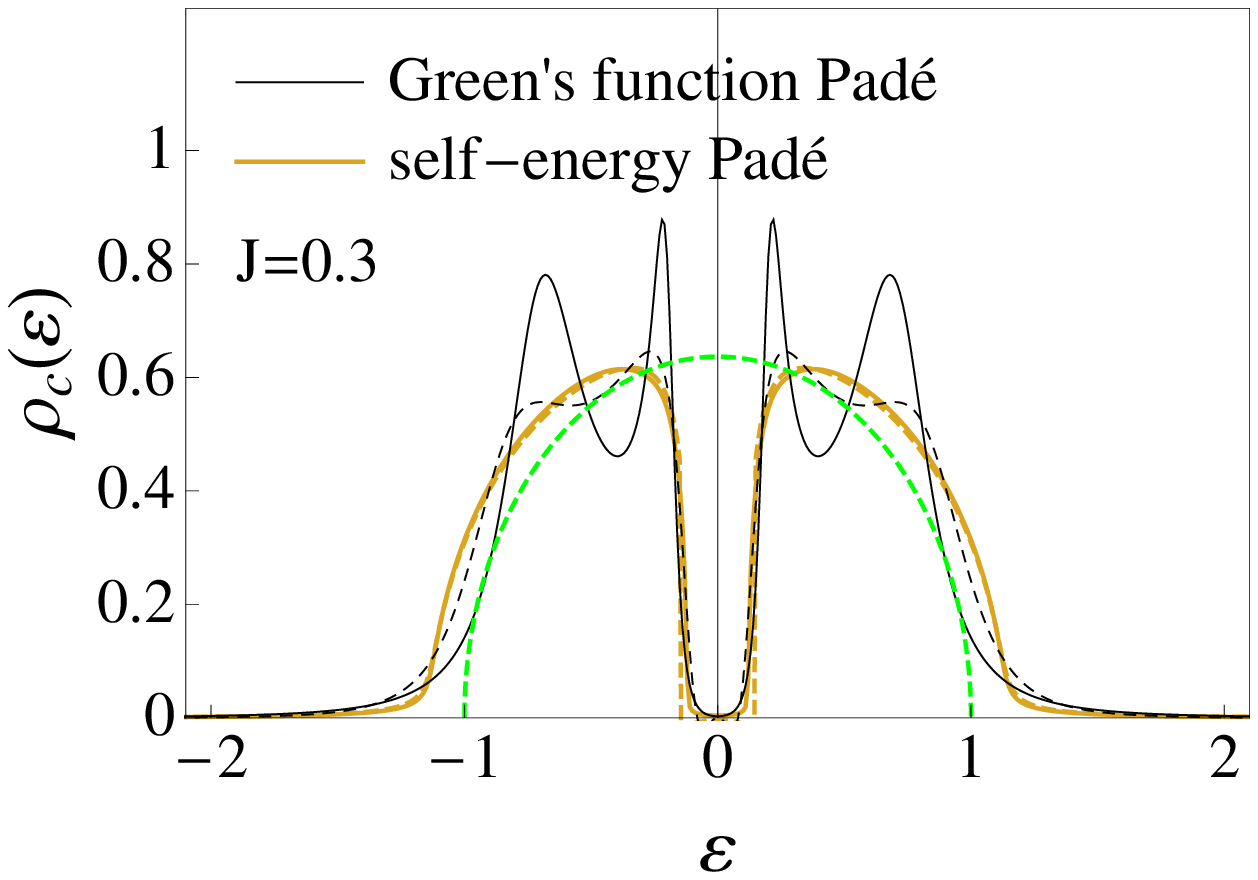}
\caption{Spectral densities at real energies for the zero-gap KLM obtained by the Green's function and self-energy Pad\'e approaches. 
The parameter values are chosen as $J=0.5$, $\beta=100$, $t=1$ ({\sl left}) and $t=0.5$ ({\sl center}).
In the {\sl right part} we show example also for the ordinary Kondo model with semicircular bare conduction electron DOS for parameter values $J=0.3$, $\beta=100$.
{\sl Continuous and dashed lines} correspond to spin-up and spin-down components, respectively. 
}
\label{figure-pade-kondo-Gc}
\end{figure}

Last, we consider a zero-gap Kondo lattice model with bare conduction electron DOS that linearly vanishes at the Fermi level and solve it numerically\cite{note-ctqmc} within DMFT.
This model corresponds to the lattice version of the pseudogap Kondo model\cite{Withoff,  Chen, Fritz_2004, Bulla1, Buxton, Vojta, Vojta_Fritz, Bulla_Glossop, Ingensent} with exponent $r=1$ which might have relevance for Kondo physics in graphene\cite{Fritz}.
Since DMFT is an infinite dimensional theory, we
take the infinite dimensional extension of the two-dimensional honeycomb and the three-dimensional diamond lattice as DOS, which is given as
\begin{eqnarray}
\rho(\omega) = \frac{|\omega|}{2t^2D} {\rm e}^{-\omega^2/(2t^2D)},
\label{eq-honeyrho0}
\end{eqnarray}
where we take the bandwidth $D=1$ as the unit of energy. 
This DOS has an extra parameter $t$ compared to the well-studied case $\rho(\omega) \sim |\omega|^{r}$, which controls the slope of the linear dispersion, i.e. the density of conduction electrons being around the Fermi level.

First, we compare spectral densities calculated at real energies by the Green's function and self-energy Pad\'e approaches shown in Fig.~\ref{figure-pade-kondo-Gc} at selected values of the Kondo coupling $J$, $t$, and temperature in the Kondo insulating phase. The insulating gap is smaller for $t=1$ than for $t=0.5$ since the slope of the linear dispersion, which is given as $1/(2t^2D)$ from Eq.~(\ref{eq-honeyrho0}), is smaller for the $t=1$ case. As a comparison, we show also example for the ordinary Kondo model with semicircular, i.e finite conduction electron DOS at the Fermi level.
We find in all cases again that the modified Pad\'e approach is more stable than the Green's function Pad\'e approach, which is indicated by the almost complete overlap of the 
spin-up and spin-down components.
Although both methods give similar result near the Fermi level, the Green's function Pad\'e approach becomes uncertain at larger energies.

\begin{figure}
\centering
\includegraphics[width= 0.45\hsize]{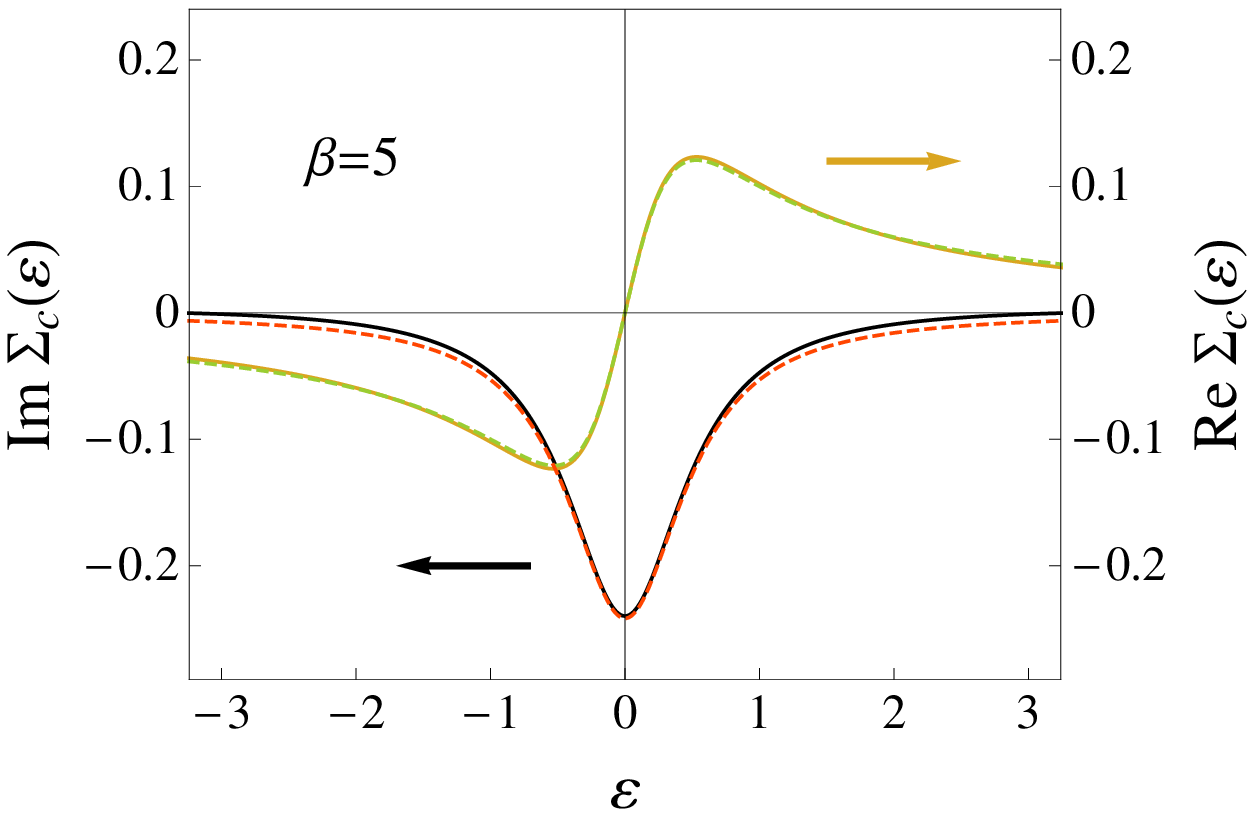}\hspace*{5mm}
\includegraphics[width= 0.45\hsize]{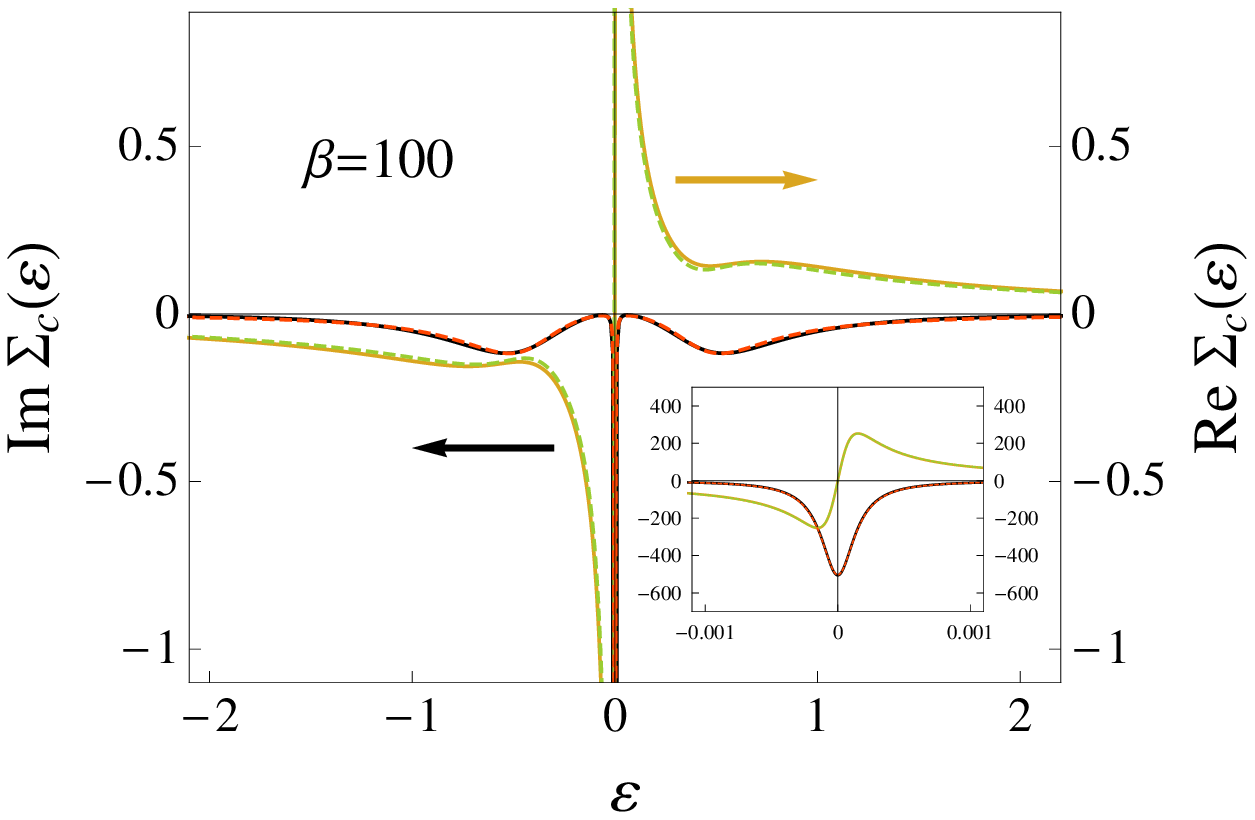}
\caption{Real energy dependence of the conduction electron self-energy for the zero-gap KLM at temperatures $\beta=5$ ({\sl left}) and $\beta=100$ ({\sl right}). The {\sl inset} shows the enlarged spectrum at the small energy range for $\beta=100$.
{\sl Green and red dashed lines} are fitting results discussed in the main text.
}
\label{figure-selfenergy-kondo-1}
\end{figure}

Figure~\ref{figure-selfenergy-kondo-1} shows the analytically continued self-energy as a function of real energy both in the high- and low-temperature phases.
The imaginary part of the self-energy shows a simple one-peak structure around the Fermi level at high temperatures, while two additional peaks evolve at higher energies with DOS origin as the temperature is decreased.
Furthermore, the central peak narrows with decreasing temperature which indicates the continuous evolution of a small energy scale responsible for the Kondo effect.
We find again that the smooth energy dependence of $\Sigma$ can be described well by the phenomenological form
\begin{eqnarray}
\Sigma^{(\rm approx)}_{c,{\rm low-T}}(\varepsilon) = \sigma(0,1,1,\varepsilon)  + \sigma(\delta, z_1, y_1,\varepsilon)
\label{eq-selfenergy-approx-II}
\end{eqnarray}
in the low-temperature phase, and as
\begin{eqnarray}
\Sigma^{(\rm approx)}_{c,{\rm high-T}}(\varepsilon) = \sigma(0,1,1,\varepsilon) = \frac{\Gamma}{\varepsilon + i\Delta}
\label{eq-selfenergy-approx-I}
\end{eqnarray}
in the high-temperature phase, where the function $\sigma$ is defined in Eq.~(\ref{eq-sigmafv}).
These fits are also included in Fig.~\ref{figure-selfenergy-kondo-1}.
We argue that the rational function nature of expressions~(\ref{eq-selfenergy-approx-II}) and (\ref{eq-selfenergy-approx-I}) is responsible for the success of the self-energy Pad\'e method similarly to the case of the Hubbard model.

\begin{figure}
\centering
\includegraphics[width= 0.45\hsize]{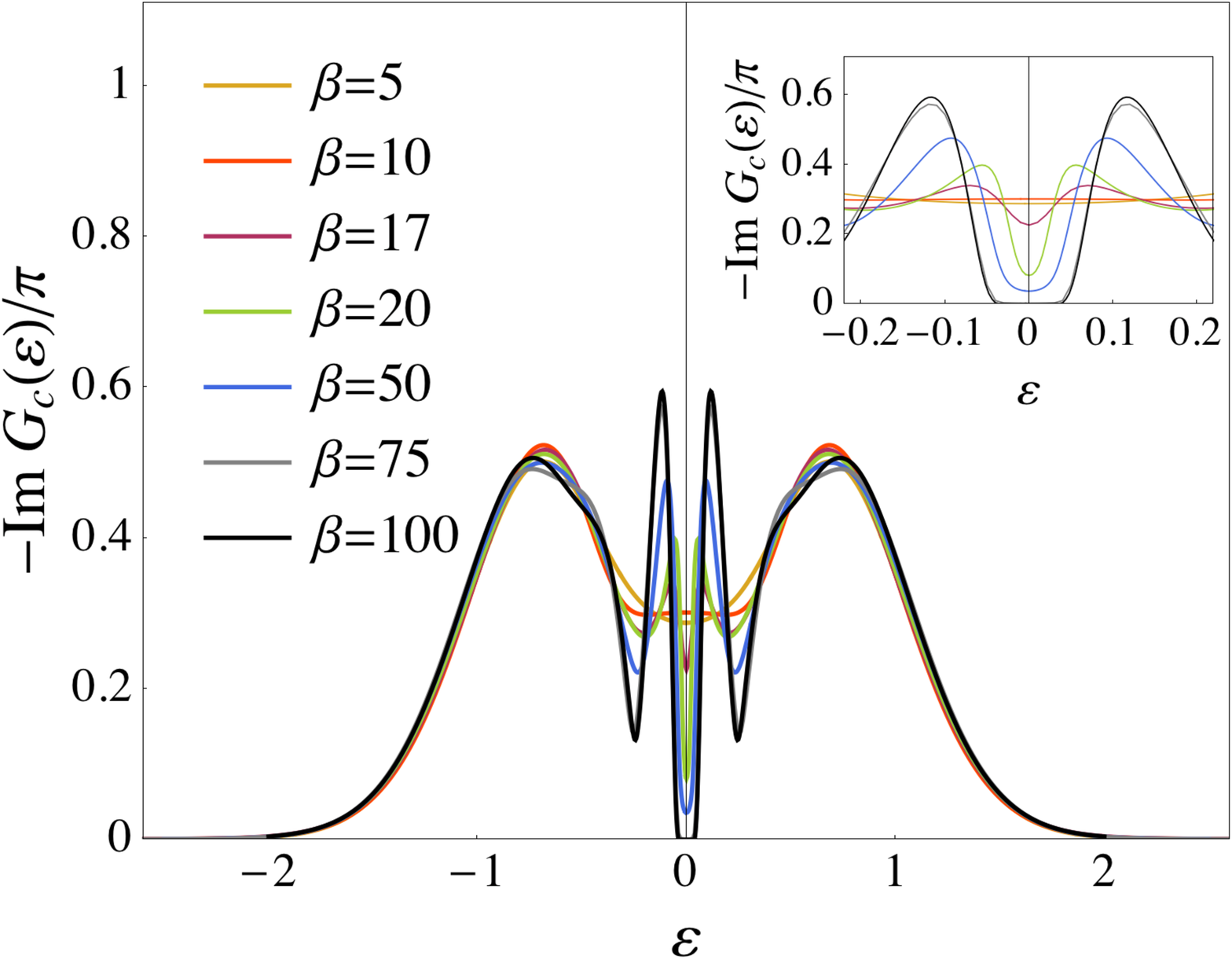}\hspace*{5mm}
\includegraphics[width= 0.45\hsize]{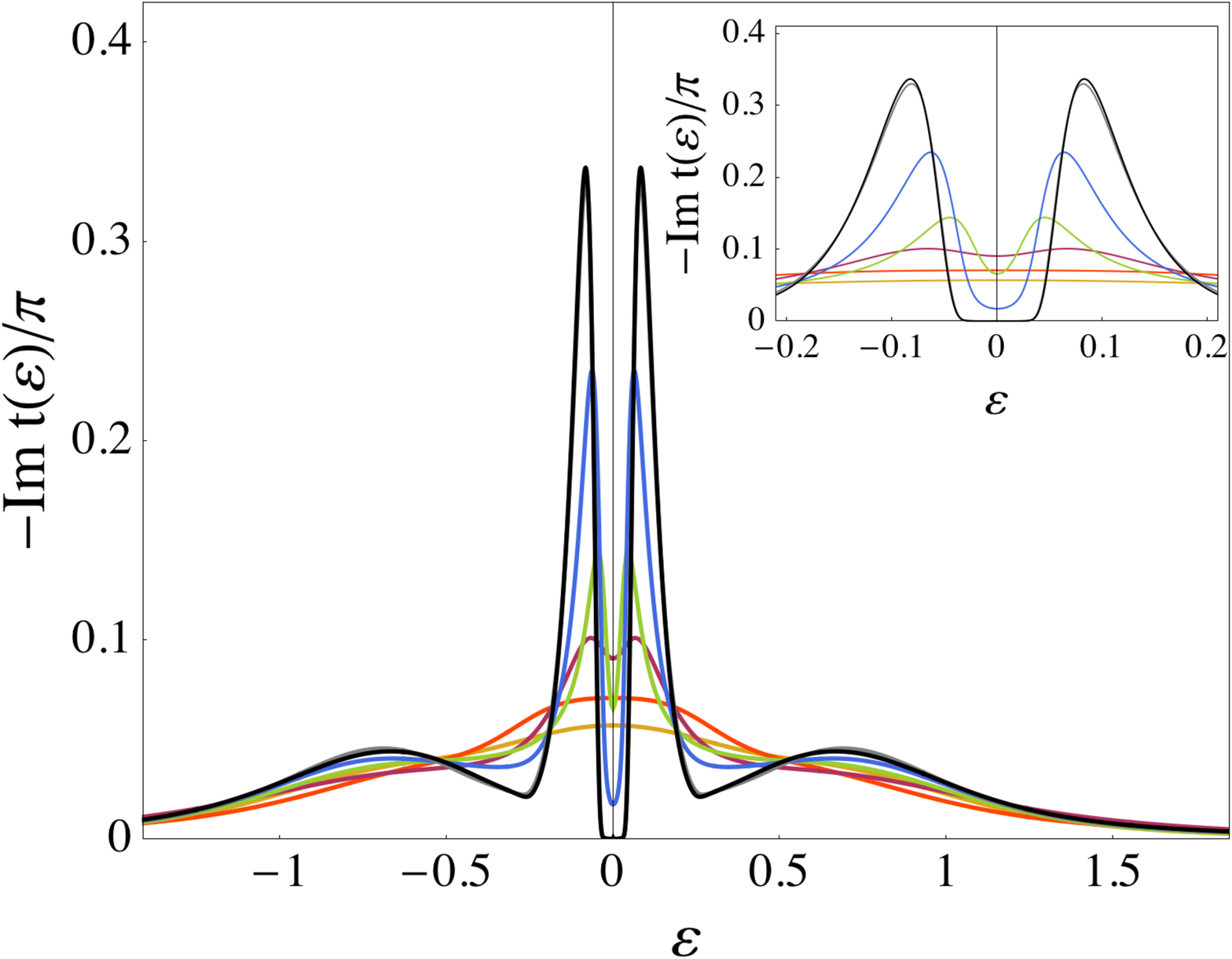}
\caption{
Conduction electron DOS ({\sl left}) and $t$-matrix ({\sl right}) for the zero-gap KLM at real energies with different values of temperature. The parameter values are chosen as $t=0.5$, $J=0.3$.
The {\sl insets} are enlarged view around the Fermi level.
}
\label{figure-peaks}
\end{figure}

We note that the conduction electron self-energy has different origin in the Kondo lattice and in the Hubbard model, although the DMFT equations are the same in these two cases.
In the case of the Kondo  model the self-energy arises from the impurity scattering while in the Hubbard model it emerges because of the Coulomb interaction. 

\begin{figure}
\centering
\includegraphics[width= 0.45\hsize]{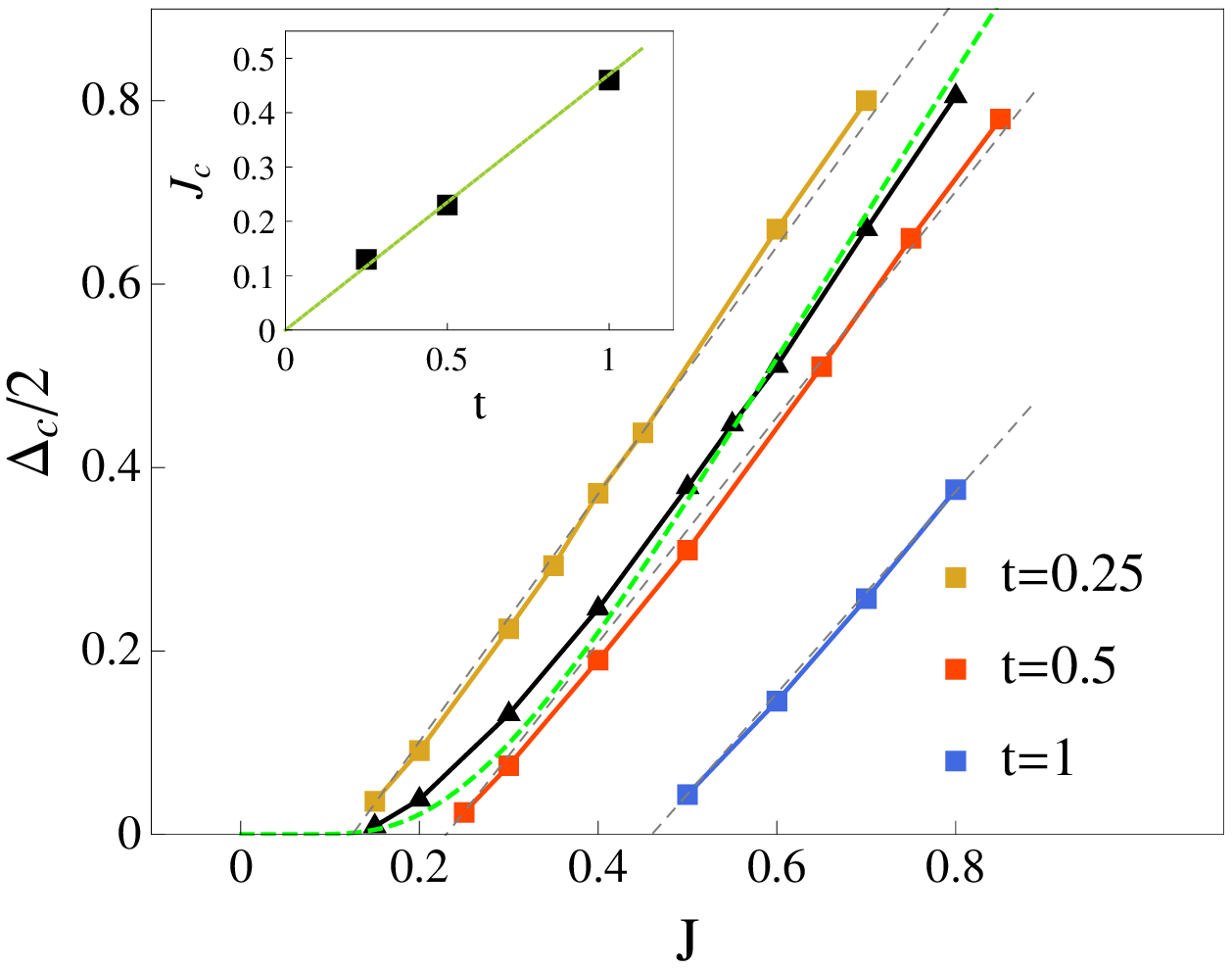}\hspace*{5mm}
\includegraphics[width= 0.45\hsize]{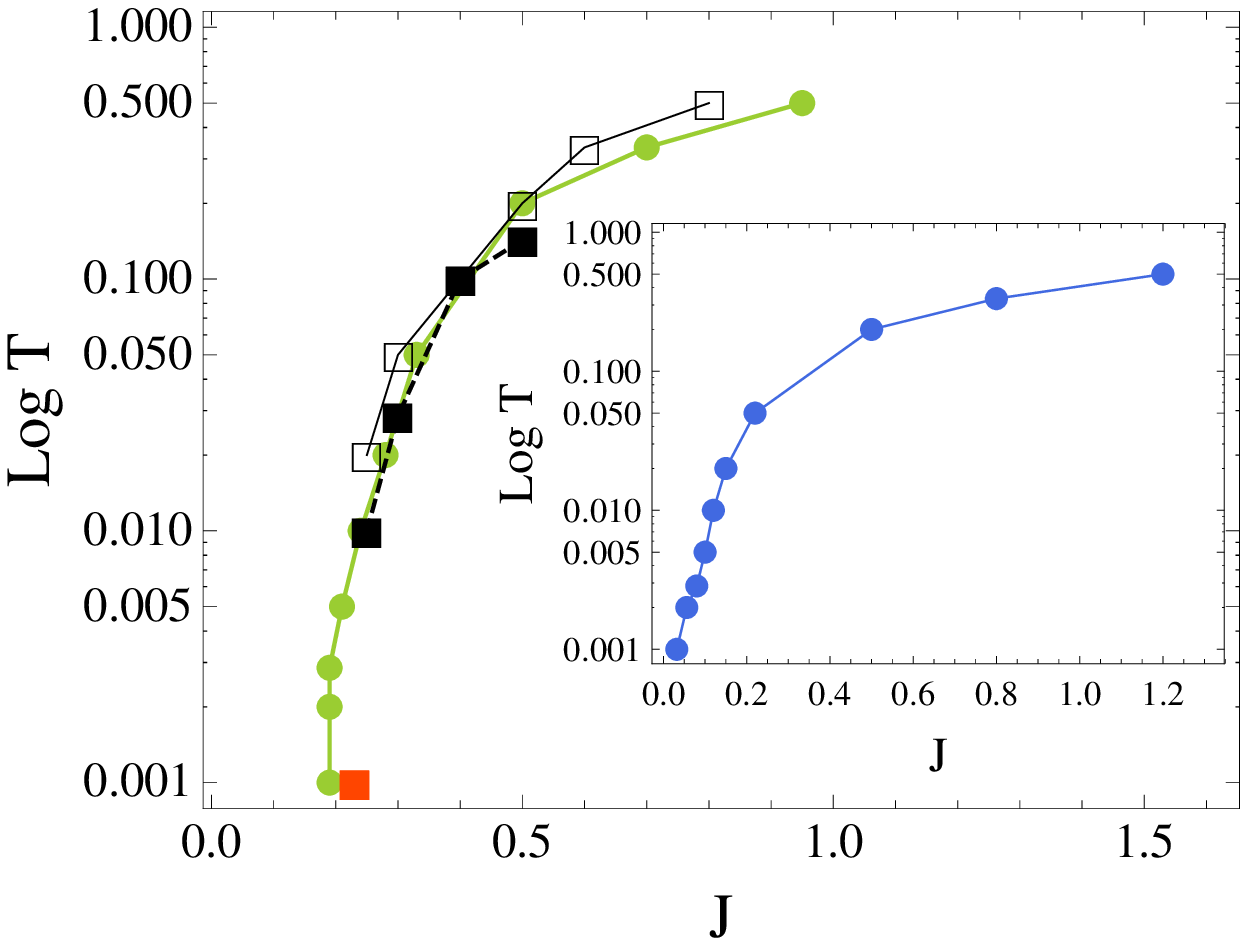}
\caption{{\sl Left:} Insulating gap ($\Delta_c$) in the spectral density as a function of Kondo coupling $J$ 
for different values of parameter $t$. As comparison, result for the ordinary Kondo model with semicircular bare conduction electron DOS is also shown by {\sl black} symbols together with a fit by the scaling result given in Eq.~(\ref{eq-scalingres}). The {\sl inset} shows the critical coupling $J_{c}$ for different $t$ values.  
{\sl Right:} 
Coherence temperature $T_{c}$ as a function of $J$ for $t=0.5$ shown by {\sl filled black boxes}.
{\sl Open black boxes} correspond to the temperature values where the insulating gap appears in the spectrum.
The ground state critical Kondo coupling $J_{c} \approx 0.23$ is shown by {\sl red box}.
{\sl Green dot} symbols show the position of inflextion points as $\{J_{\rm infl},T\}$ for the zero-gap KLM, while the  {\sl inset} shows the result for the semicircular case. 
}
\label{figure-criticalJ}
\end{figure}

Figure~\ref{figure-peaks} shows the temperature dependence of the conduction electron DOS  and $t$-matrix  at real energies in the Kondo insulating phase, i.e.
where an energy gap develops as the temperature is decreased.
The $J$-dependence of this insulating gap, $\Delta_{c}$, is shown in the left part of Fig.~\ref{figure-criticalJ} for different values of the DOS parameter $t$. 
The gap $\Delta_c$ is measured at the half-peak value of $\rho_{c}(\varepsilon)$ since it becomes almost temperature independent at low temperatures.
We find a linear dependence of $\Delta_c$ on the Kondo coupling $J$ for all values of parameter $t$, which is in contrast to the normal Kondo case with finite DOS at the Fermi level. 
Namely, we show also the gap for the Kondo lattice with semicircular bare conduction electron DOS in Fig.~\ref{figure-criticalJ} as comparison, which is in agreement with the scaling formula for the Kondo temperature as
\begin{eqnarray}
\Delta_{c} \sim T_{K}^{(\rm scaling)} = D (N J \rho_{c})^{1/N} {\rm e}^{-1/(N J \rho_{c})}
\label{eq-scalingres}
\end{eqnarray}
obtained for the $N$-fold degenerate Coqblin-Schrieffer model\cite{hewson} with $D$ being the band half-width and $\rho_{0}$ being constant conduction electron DOS.
In accord with the scaling formula~(\ref{eq-scalingres}) Kondo insulating state appears for any value of the Kondo coupling $J$ at low enough temperature due to the finite DOS at the Fermi level. 
In contrast, there is a critical value $J_{c}$ of the Kondo coupling for the formation of the low-temperature insulating phase for the zero-gap KLM as it is apparent from Fig.~\ref{figure-criticalJ}.
The critical values $J_{c}$  are shown in the inset of the left part of Fig.~\ref{figure-criticalJ} for different $t$ values, which are derived from the linear fit to the gap values $\Delta_{c}$.
In the limit of $t \rightarrow 0$ the critical coupling  $J_{c}$ tends to zero, i.e. the case of finite DOS at the Fermi level is recovered since the slope of the dispersion is infinite in the $t=0$ limit.

\begin{figure}
\centering
\includegraphics[width= 0.45\hsize]{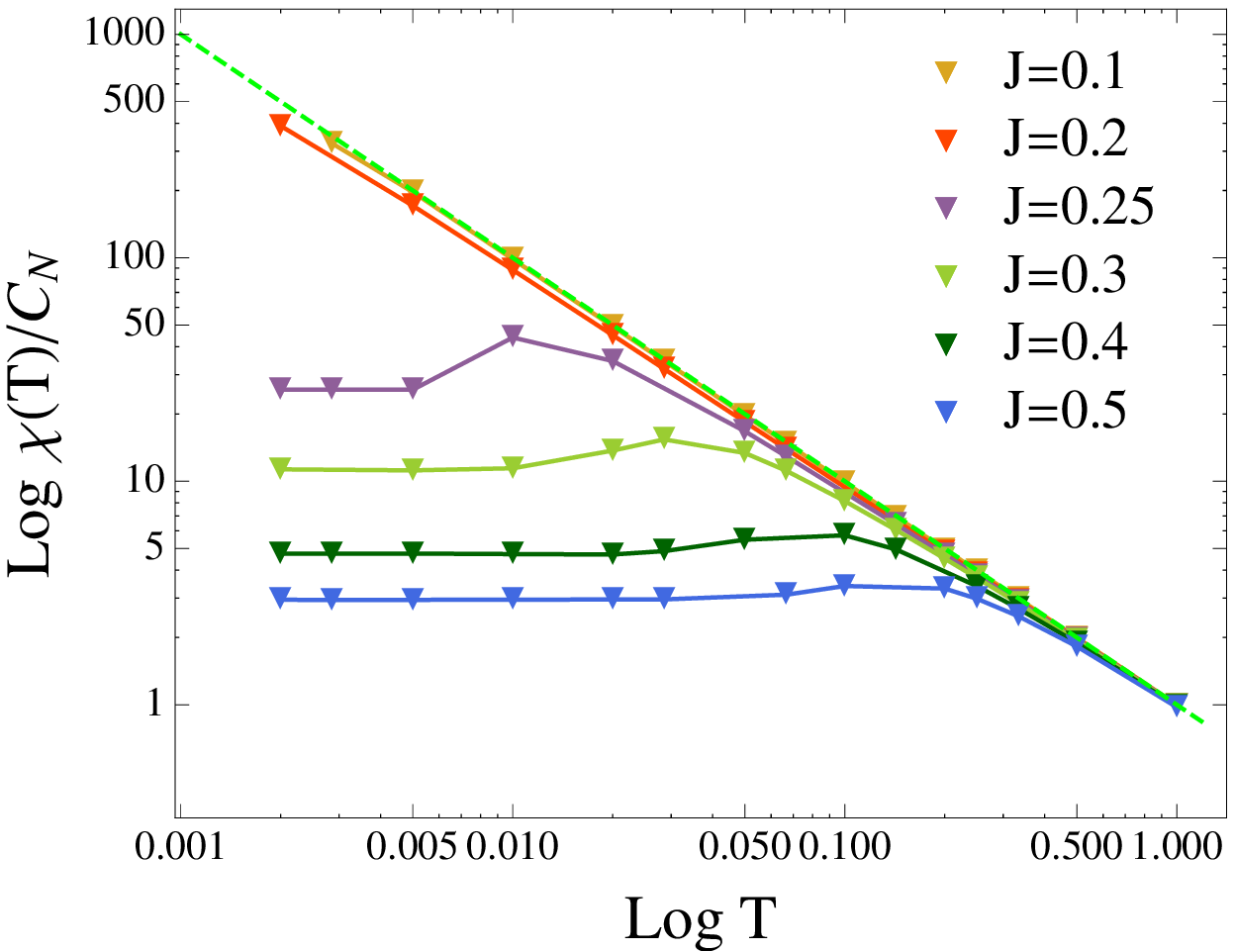}\hspace*{5mm}
\includegraphics[width= 0.45\hsize]{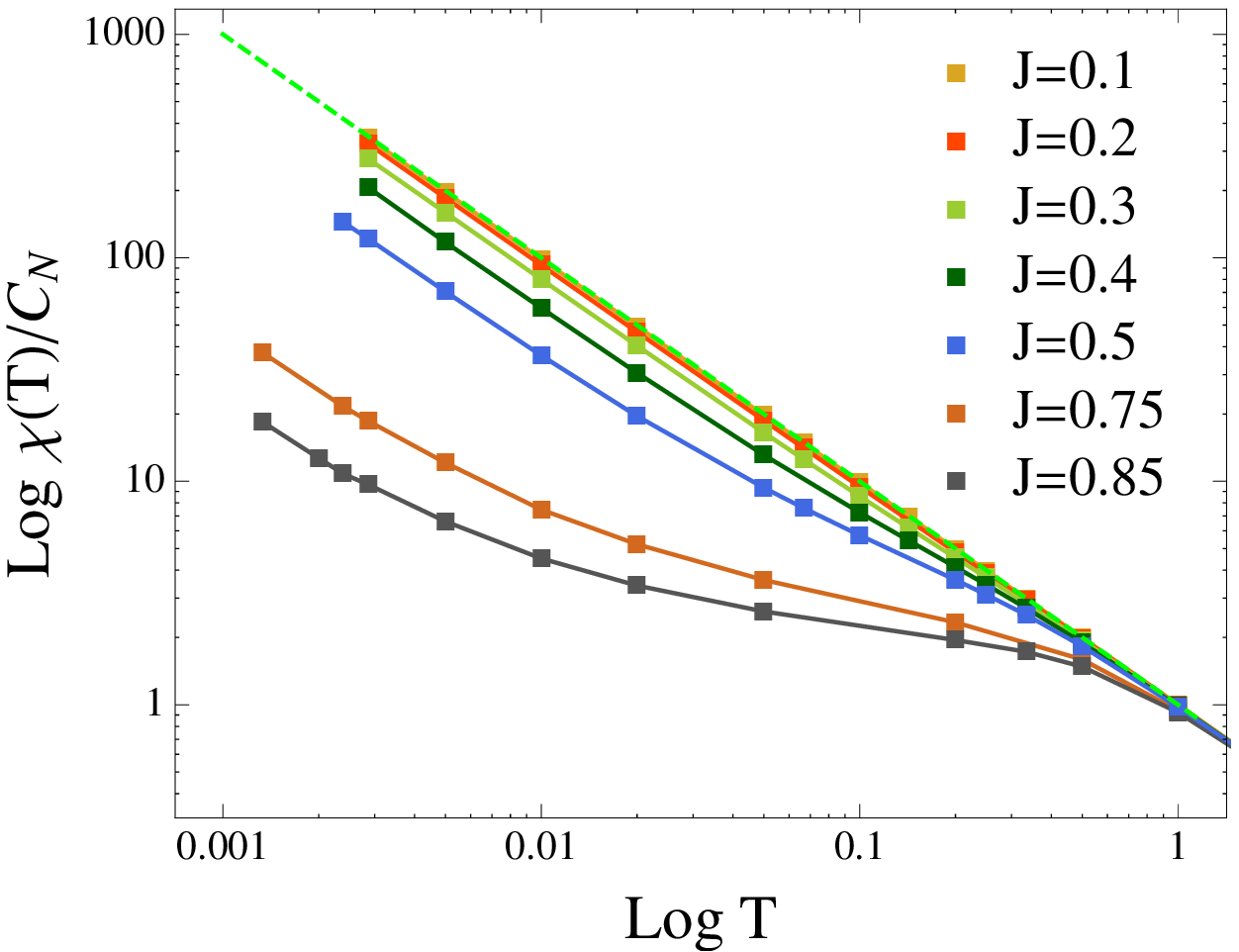}
\caption{Temperature dependence of the static local susceptibility with different values of Kondo coupling $J$ 
for the Kondo lattice ({\sl left}) with $t=0.5$ and for the corresponding Kondo impurity problem ({\sl right}).
The {\sl dashed green line} shows $1/T$ dependence.
}
\label{figure-susceptibility}
\end{figure}

The temperature dependence of the static local susceptibility is shown in the left part of Fig.~\ref{figure-susceptibility} for different values of the Kondo coupling $J$. 
For $J<J_{c}$ the susceptibility increases with decreasing temperature approximately as $\sim 1/T$,
while it saturates at low temperatures for $J \ge J_{c}$.
The critical Kondo coupling is estimated as $J_{c} \approx 0.23$ for $t=0.5$ in our numerical calculations (see left part of Fig.~\ref{figure-criticalJ}).
We define the Kondo temperature from the susceptibility as $T_{\rm K}^{-1} = \chi(T \rightarrow 0)/C_{N}$, where $C_{N}$ is the Curie constant.
Beside $T_{\rm K}$, another energy scale $T_{\rm c}$ can be identified which is the temperature
where the susceptibility has maximum in its $T$-dependence.
Below this temperature, coherence between the impurities appear in a Kondo lattice by coherent scattering off the Kondo singlets, which is shown by the behavior $\chi \sim \, {\rm constant}$ for $T<T_{\rm c}$.
The $J$-dependence of the coherence temperature $T_{\rm c}$ is shown in the right part of Fig.~\ref{figure-criticalJ}.

We expect that coherence effects are reflected in further thermodynamic and dynamic quantities as well. 
Actually we find that the temperature values where the insulating gap appears in the spectral density at a given value of $J$ coincide well with $T_{\rm c}$ as it is shown in Fig.~\ref{figure-criticalJ}.
Furthermore, if we obtain the inflextion points\cite{note-2} $\{J_{\rm infl},T_{\rm infl}\}$ of the wave-function renormalization factor $z=(1-\partial {\rm Re} \Sigma_{c}(\varepsilon)/\partial \varepsilon |_{\varepsilon=0})^{-1}$ from its $J$- and $T$-dependence, and plot $T_{\rm infl}$ as a function of $J$, we find that this curve also coincides  with the previous ones as it can be seen in Fig.~\ref{figure-criticalJ}. 
All these quantities are finite only above a critical value of the Kondo coupling which matches well with the critical value $J_{c} \approx 0.23$ obtained previously from the vanishing of the insulating gap $\Delta_{c}$.
On the other hand, there is no such critical value for the case of the normal Kondo lattice model as it is shown in the inset of the right part of Fig.~\ref{figure-criticalJ}.

We note that in contrast to the lattice case, there is no Kondo effect in the corresponding zero-gap Kondo impurity problem even for infinitely strong Kondo coupling, which is illustrated by the temperature dependence of the static local susceptibility shown in the right part of Fig.~\ref{figure-susceptibility}. Namely, the susceptibility follows $1/T$ dependence with a $J$-dependent constant as the temperature is lowered for even large values of $J$ (e.g. $J=0.85$ in Fig.~\ref{figure-susceptibility}), which reflects that the impurity is in local moment state without screening. Kondo screening of the local moment in the impurity case is possible in particle-hole asymmetric case which can be achieved by the introduction of non-zero chemical potential, for example, or including potential scattering.

\begin{figure}
\centering
\includegraphics[width= 0.45\hsize]{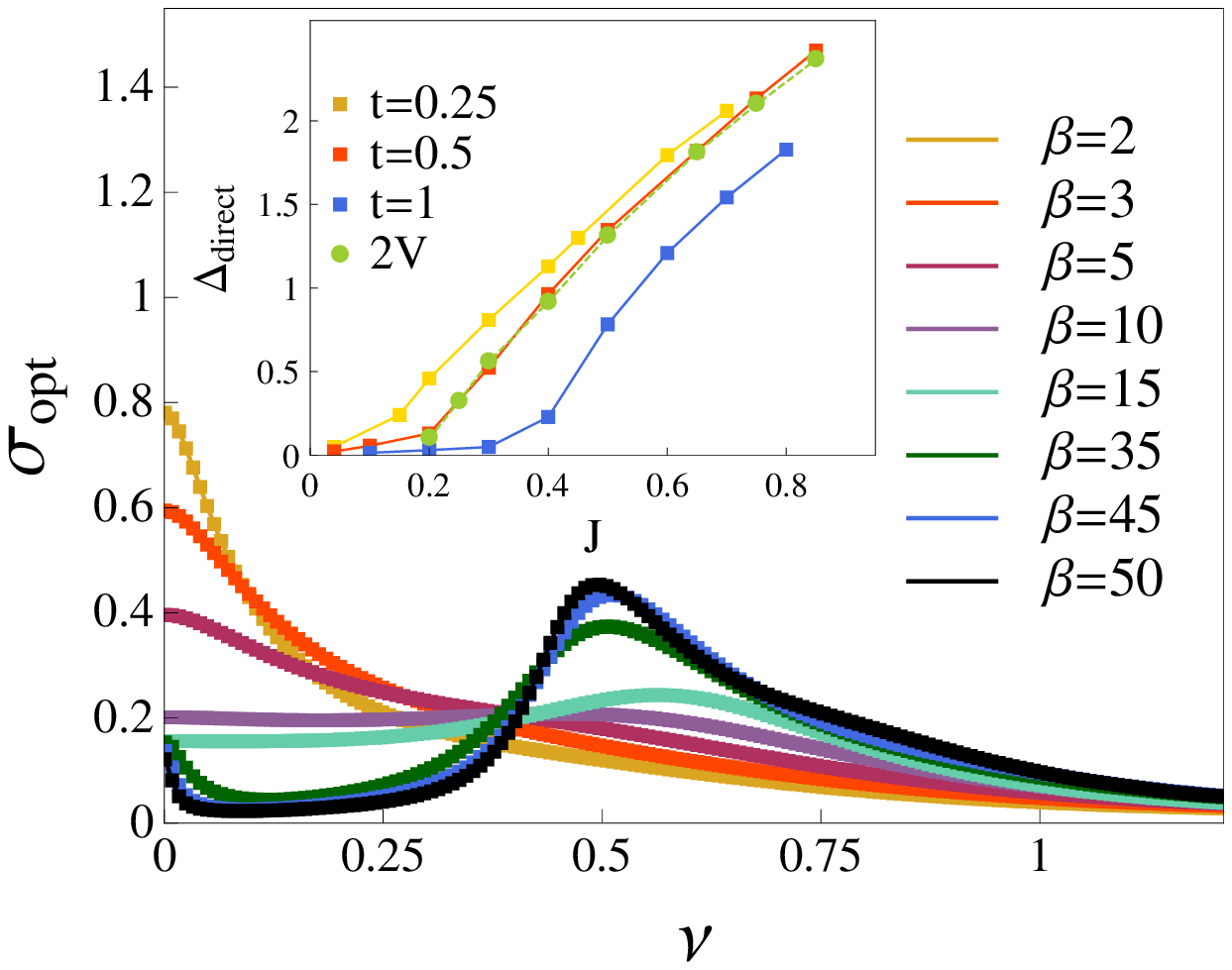}\hspace*{5mm}
\includegraphics[width= 0.45\hsize]{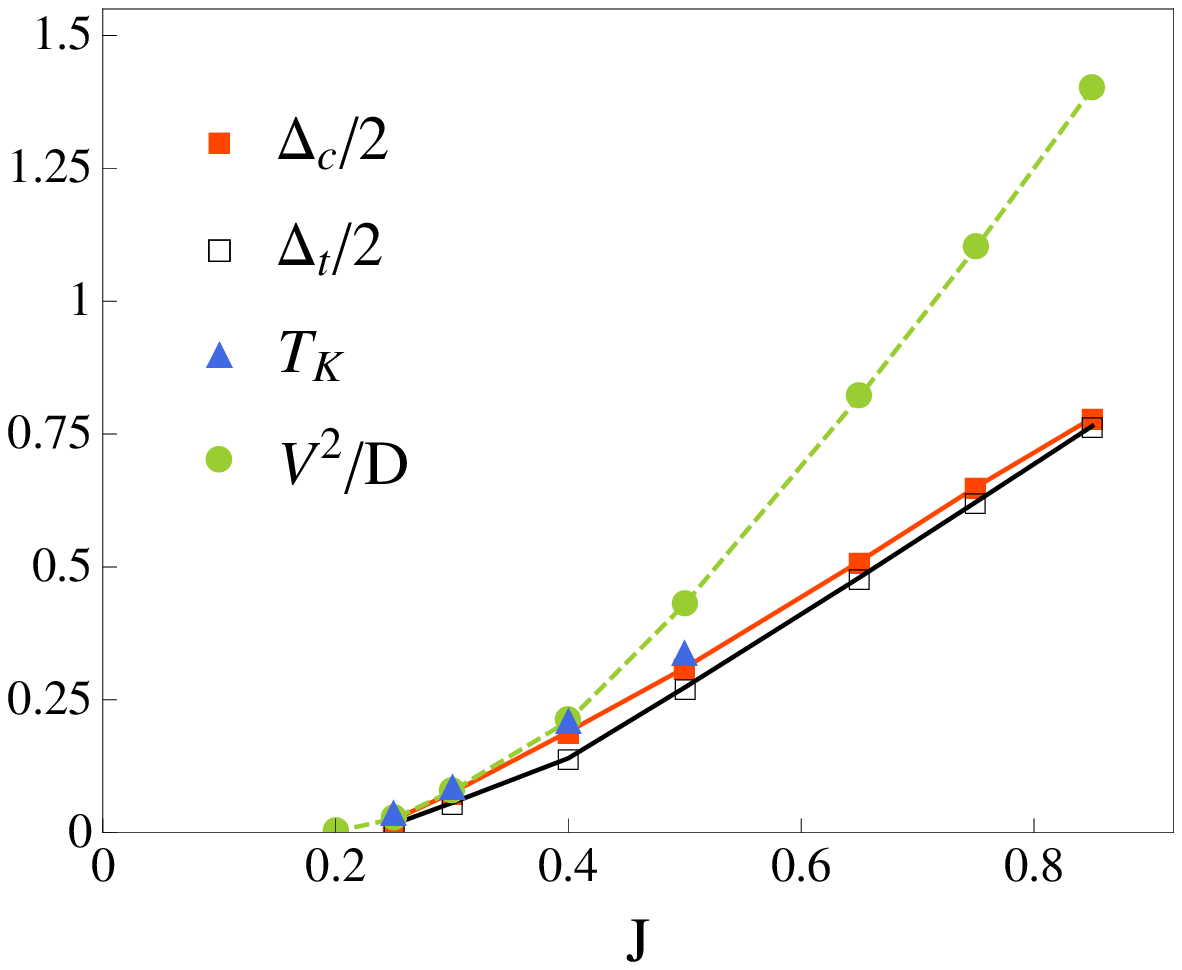}
\caption{ 
{\sl Left:} Optical conductivity of the zero-gap Kondo lattice with $t=0.5$ as a function of frequency with decreasing temperature indicated with darker colors. 
The {\sl inset} shows the direct gap values for different Kondo couplings $J$ and parameter $t$ obtained from the optical conductivity.
{\sl Right:} Indirect gap values $\Delta_{c}/2$ and $\Delta_{t}/2$ obtained from the energy dependence of the conduction electron DOS and from the $t$-matrix, respectively.  
{\sl Dashed green lines} show fit of the numerical data explained in the main text.
}
\label{figure-gaps}
\end{figure}

Right part of Fig.~\ref{figure-gaps} shows the temperature evolution of the optical conductivity as a function of frequency, where the optical conductivity is calculated as \cite{pruschke}
 \begin{eqnarray}
{\rm Re}\, \sigma(\nu) &=&  \int d\varepsilon  
 \int_{-\infty}^{\infty} d\omega^{\prime}
\rho_{0}(\varepsilon) \rho_{c}(\varepsilon,\omega^{\prime}) \rho_{c}(\varepsilon,\omega^{\prime}+\nu)
 \frac{f(\omega^{\prime}) - f(\omega^{\prime}+\nu)}{\nu}
\label{sigma8}
\end{eqnarray}
with $f(\omega) = 1/(1+{\rm exp}(\beta \omega ))$ being the Fermi function.
We can observe that a peak emerges at a finite frequency as the temperature is decreased which indicates the presence of a direct gap in the energy spectrum. 
The positions of these peaks, that we identify as direct gap, are shown in the inset as a function of Kondo coupling $J$ for different values of parameter $t$.
Beside this direct gap, we identify the insulating gap $\Delta_{\rm c}$ from dynamics as indirect gap.
Right part of Fig.~\ref{figure-gaps} shows this indirect gap together with $\Delta_t$ obtained from the $t$-matrix, and the Kondo temperature $T_{\rm K}$ as well.

In order to interpret the direct and indirect gaps as energy scales, we recall the non-interacting Anderson lattice model.
From the exact energy spectrum given in Eq.~(\ref{and-band-e}) for the ALM, these two energy gaps can be defined. 
The indirect gap is written as $\Delta_{\rm indirect} = E_{+}(k \rightarrow \infty)-E_{-}(k \rightarrow -\infty) \approx 2V^2/D$ that emerges due to the hybridization in dynamical quantities such as conduction electron or local DOS.
The direct gap is expressed as $\Delta_{\rm direct} = E_{+}(k=0)-E_{-}(k=0) \approx 2V$.
We recall Eq.~(\ref{Scz}) which says $\Sigma_{c}(i\varepsilon_{n}) = V^2/(i\varepsilon_{n})$ for the self-energy in the case of the non-interacting  ALM. 
Thus, we obtain the associated term "$V^2$" for the KLM
by extracting numerically the coefficient of $1/\varepsilon_{n}$  in the self-energy $\Sigma_{c}(i\varepsilon_{n}) $ in the Kondo lattice model, and with the use of this associated term "$V^2$" we estimate the gaps as $\Delta_{\rm indirect} \sim 2V^2/D (=2V^2)$ and $\Delta_{\rm direct} \sim 2V$ which are also shown in Fig.~\ref{figure-gaps}.
We find an excellent agreement between $2V$ and the direct gap of KLM obtained from the optical conductivity data as it is shown in the inset of the left part of Fig.~\ref{figure-gaps}. For the indirect gap we expect that $\Delta_c/2 \sim V^2/D$, but instead we find from the numerical analysis that $\Delta_c/2 \sim 0.5 (V^2/D)$ for parameter value $t=0.5$ as it is shown in the right part of Fig.~\ref{figure-gaps}.
The reason for the discrepancy might be that our identification between the indirect gaps of the non-interacting ALM and KLM is just approximate and does not contain interaction effects which are expected to influence the indirect gap stronger than the direct gap.
Nevertheless, the hybridized bands picture with direct and indirect gaps seems to be held for the zero-gap Kondo lattice model as well.

\section{Summary}\label{summary}

In this paper we have presented and discussed a modified Pad\'e approach for numerical methods in which quantities like correlation functions are obtained in imaginary time. 
The modified Pad\'e approach is based on the analytic continuation of the self-energy which is then used for deriving the Green's function and $t$-matrix at real energies.
We showed that this modified, self-energy Pad\'e approach is more stable and robust against statistical errors than the conventional way with the direct analytic continuation of the Green's function, where the latter property is demonstrated within a toy model.
We found that the success of the modified Pad\'e approach lies in the fact that the self-energy has smooth real-energy dependence and this dependence can be approximated excellently by rational function, therefore, the Pad\'e approximation works well for the self-energy.
We considered also the zero-gap Kondo lattice with linearly vanishing conduction electron DOS at the Fermi level where we applied the modified Pad\'e approach, and investigated the properties of its Kondo insulating phase.
We found linear dependence of the insulating gap on the Kondo coupling $J$ with a finite critical value $J_{c}$ above which Kondo insulating state appears.
Besides the Kondo temperature $T_{\rm K}$, we identified the coherence temperature  $T_{\rm c}$ as well from thermodynamic and dynamic quantities below which coherence between the impurities appear.
Finally, we have identified two additional energy scales from dynamic and thermodynamic properties, that are associated as a direct and an indirect gap in a band hybridization picture.

\section*{Acknowledgment}

The author would like to acknowledge helpful advices and discussion with Professor Yoshio Kuramoto.
The author acknowledges the Bolyai Program of the Hungarian Academy of Sciences and the Hungarian Scientific Research Funds No. K124176.

\appendix

\section{Pad\'e approximation}\label{sec-app1}

The principal problem in analytic continuation by using Pad\'e approximation is to find a rational function $C_{N}(n)$ which approximates the function $G(n)$ at arbitrary real energies $n$, $G(n) \sim C_{N}(n)$, on condition that we know the function $G$ at $N$ Matsubara frequency points $n_{1}, n_{2}, ... ,n_{N}$.
The rational function $C_{N}(n)$ is expressed as
\begin{eqnarray}
C_{N}(n) = \frac{A_{N}(n)}{B_{N}(n)} = \frac{a_1}{1+\frac{a_2(n-n_1)}{1+\frac{a_3(n-n_2)}{1+...}}},\label{pade1}
\end{eqnarray}
where the coefficients $a_i$ are determined from the condition $C_{N}(n_i) = G(n_i)$ for $i=1,...,N$.

The simplest way for obtaining the coefficients $a_i$ is the use functions $g$ as $a_i = g_{i}(n_i)$, where $g_{r}(n)$ can be obtained by the following recursion formula:
\begin{eqnarray}
&&{\rm for} \,\, r=1: \,\, g_{1}(n_{i}) = G(n_i), \\
&&{\rm for} \,\, r>1: \,\, g_{r}(n) = \frac{g_{r-1}(n_{r-1}) - g_{r-1}(n)}{(n - n_{r-1})g_{r-1}(n)},
\end{eqnarray}
which gives the functions $A_{N}(n)$ and $B_{N}(n)$ in Eq.~(\ref{pade1}) as
\begin{eqnarray}
A_{q+1}(n) &=& A_{q}(n) + (n - n_q) a_{q+1}A_{q-1}(n),\\
B_{q+1}(n) &=& B_{q}(n) + (n - n_q) a_{q+1}B_{q-1}(n)
\end{eqnarray}
with the starting conditions $A_{0}=0$, $A_1 = a_1$, $B_0=B_1=1$.

\section{Non-interacting Anderson lattice model}\label{sec-app2}

In the DMFT treatment of the Anderson lattice model given in Eq.~(\ref{and-lattice-model}) the site-diagonal Green's function is calculated as
\begin{eqnarray}
\overline{G}_{f}(i\varepsilon_{n}) = \int d\omega \rho(\omega) \left( i\varepsilon_{n} - \varepsilon_{f} - \Sigma_{f}(i\varepsilon_{n}) - \frac{V^2}{i\varepsilon_{n} - \omega + \mu} \right)^{-1}
\label{eq-greenf} 
\end{eqnarray}
in the Matsubara frequency domain with $\varepsilon_{n} = (2n+1)\pi T$,
and $\rho(\omega)$ is the bare conduction electron DOS of the conduction electrons.
We quote also the results for the conduction electron self-energy and Green's function, which read as
\begin{eqnarray}
\Sigma_{c}(i\varepsilon_{n}) = \frac{V^2}{i\varepsilon_{n} - \varepsilon_{f} + \Sigma_{f}(i\varepsilon_{n})} ,
\label{eq-sigmac}
\end{eqnarray}
and
\begin{eqnarray}
\overline{G}_{c}(i\varepsilon_{n}) = \int d\omega \rho(\omega) \left( i\varepsilon_{n} - \omega  + \mu - \Sigma_{c}(i\varepsilon_{n}) \right)^{-1}.
\label{eq-greenc} 
\end{eqnarray}

In the non-interacting case, the local self-energy vanishes, i.e. $\Sigma_{f}=0$, and thus from Eqs.~(\ref{eq-greenf}), (\ref{eq-sigmac}) and  (\ref{eq-greenc}) the local and conduction electron Green's functions can be evaluated exactly in the Matsubara frequency domain as
\begin{eqnarray}
G_{c}({\mathbf k}, i\varepsilon_{n}) &=& \left(i\varepsilon_{n} - \varepsilon_{\mathbf k} - \frac{V^2}{ i\varepsilon_{n}- \varepsilon_{f}} \right)^{-1},\\
G_{f}({\mathbf k}, i\varepsilon_{n}) &=& \left(i\varepsilon_{n} - \varepsilon_{f} - \frac{V^2}{i\varepsilon_{n} - \varepsilon_{\mathbf k}} \right)^{-1}.
\end{eqnarray}
The conduction and local electron DOS are obtained at real energies as
\begin{eqnarray}
\rho_{c}(\varepsilon) &=&  -\frac{1}{\pi} {\rm Im}\, \sum_{\mathbf k} G_{c}({\mathbf k}, \varepsilon) =  \rho\left(\varepsilon -  \frac{V^2}{ \varepsilon- \varepsilon_{f} } \right),
\label{eq-rhoc-and-exact}\\
\rho_{f}(\varepsilon) &=& -\frac{1}{\pi} {\rm Im}\, \sum_{\mathbf k} G_{f}({\mathbf k}, \varepsilon) =  
 \frac{V^2}{ (\varepsilon- \varepsilon_{f})^2} \rho\left(\varepsilon -  \frac{V^2}{ \varepsilon- \varepsilon_{f} } \right) ,
 \label{eq-rhof-and-exact}
\end{eqnarray}
after performing the ${\mathbf k}$-summations, which can be expressed by the self-energy as well. Namely, $\rho_{c}(\varepsilon)$ was already expressed in Eq.~(\ref{eq-greenc3}), and the local DOS is expressed as
\begin{eqnarray}
\rho_{f}(\varepsilon) 
= -\frac{1}{\pi} {\rm Im}\, \int d\omega  \rho(\omega) \left( \frac{V^2}{\Sigma_{c}(\varepsilon)} - \frac{V^2}{\varepsilon - \omega} \right)^{-1},
\label{eq-sigmacmatsanalytic}
\end{eqnarray}
where we used Eqs.~(\ref{eq-greenf}) and (\ref{eq-sigmac}).

The energy of the hybridized bands between the local and the conduction electron states in the non-interacting Anderson lattice model is expressed as
\begin{eqnarray}
E_{\pm} = \frac{\varepsilon_{\mathbf k} - \varepsilon_{f}}{2} \pm \sqrt{ \left(  \frac{\varepsilon_{\mathbf k} - \varepsilon_{f}}{2} \right)^2 + V^2}.
\label{and-band-e}
\end{eqnarray}

\section{Pad\'e approximant for the self-energy in case of the Hubbard model}\label{app-hubbard}

We calculate the Pad\'e approximant for the phenomenological self-energy both for the metallic and insulating phases given in Eqs.~(\ref{eq-selfenergy-approx-hubbard-I-matsu}) and (\ref{eq-selfenergy-approx-hubbard-II-matsu}).
From Eq.~(\ref{eq-selfenergy-approx-hubbard-II-matsu}) we calculate the Pad\'e coefficients as
\begin{eqnarray}
a_{1} = \frac{-i\Gamma}{\varepsilon_{0} + \Delta }, \,\,\,\, a_{2} = \frac{-i}{\varepsilon_{0} + \Delta}, \,\,\,\, a_{i} = 0 \,\,\, {\rm for}\,\,\, i>2
\end{eqnarray}
based on Appendix~\ref{sec-app1},
which gives the real-energy Pad\'e approximant as
\begin{eqnarray}
\sigma(0,1,1, \varepsilon) =  \frac{a_{1}}{1+ \frac{a_{2}(\varepsilon - i\varepsilon_{0})}{1+\frac{a_{3}(\varepsilon - i\varepsilon_{1})}{1+....}}  } = 
 -\frac{i\Gamma}{(\varepsilon_{0} + \Delta) } \frac{1}{\left(1  -\frac{i}{(\varepsilon_{0} + \Delta)}(\varepsilon - i\varepsilon_{0}) \right) } = \frac{\Gamma}{  \varepsilon + i \Delta },
 \label{padeap2}
\end{eqnarray}
i.e. the phenomenological form~(\ref{eq-selfenergy-approx-hubbard-II}) is  reproduced.

In the correlated metallic phase the Pad\'e coefficients $a_{i}$ are calculated for Eq.~(\ref{eq-selfenergy-approx-hubbard-I-matsu}) as
\begin{eqnarray}
a_{1}  &=&  \frac{ \varepsilon_{0} z_1\Gamma}{iy_1 \Delta \varepsilon_{0} + i(\varepsilon_{0})^2+i\delta} ,\label{eq-a1} \\
a_{2}  &=&  \frac{i(\delta-\varepsilon_{0}\varepsilon_{1})}{(y_1 \Delta \varepsilon_{0} + (\varepsilon_{0})^2+\delta)\varepsilon_{1}} , \\
a_{3}  &=& \frac{-i\delta}{(\delta - \varepsilon_{0}\varepsilon_{2} )\varepsilon_{1}} ,\\
a_{4}  &=& \frac{ i \varepsilon_{0} }{(\delta - \varepsilon_{0}\varepsilon_{2} )},\label{eq-a4} \\
a_{i}  &=& 0 \,\,\,\, {\rm for} \,\,\,\, i>4.
\end{eqnarray}
With these coefficients the analytically continuated $\sigma$ function is obtained as
\begin{eqnarray}
\sigma(\delta, z_1, y_1, \varepsilon) =  \frac{a_{1}}{1+ \frac{a_{2}(\varepsilon - i\varepsilon_{0})}{1+\frac{a_{3}(\varepsilon - i\varepsilon_{1})}{1+ a_{4}(\varepsilon - i\varepsilon_{2}) }}  } = \frac{ \varepsilon z_1\Gamma}{  i \varepsilon y_1 \Delta  + \varepsilon^2 - \delta } = \frac{ -i \varepsilon z_1\Gamma}{  \varepsilon y_1 \Delta  - i \varepsilon^2 + i \delta }.
\label{padeap1}
\end{eqnarray}
Thus the real-energy phenomenological form~(\ref{eq-selfenergy-approx-hubbard-I}) is also reproduced.


\begin{thebibliography}{99}
\bibitem{note-ctqmc}
For numerical calculations, we use the hybridization-expansion version of the continuous-time quantum Monte Carlo method\cite{Werner1,Werner2, otsuki}  in this paper.
\bibitem{note-3}
We note that although the DMFT solution of the Hubbard model contains this Mott insulating state as well, it has finite entropy in the ground state which should be removed either by quantum fluctuations or magnetic orders beyond the scope of the DMFT.
\bibitem{note-2}
{The inflextion point can be determined by looking at the locations of the minimum of the $J$-derivative of the wave-function renormalization factor.}
\end{thebibliography}
\end{document}